\documentclass{aa}
\usepackage[varg]{txfonts}

\usepackage{natbib}
\bibpunct{(}{)}{;}{a}{}{,} 

\usepackage{graphicx}

\begin{document}

   \title{Multi-point study of the energy release and transport in the \\28 March 2022, M4-flare using STIX, EUI, and AIA during \\the first Solar Orbiter nominal mission perihelion}


   \author{Stefan Purkhart\inst{1},
          Astrid M. Veronig\inst{1}\fnmsep\inst{2},
          Ewan C. M. Dickson\inst{1},
          Andrea Francesco Battaglia\inst{3}\fnmsep\inst{4},
          Säm Krucker\inst{3}\fnmsep\inst{5},
          Robert Jarolim\inst{1},
          Bernhard Kliem\inst{6},
          Karin Dissauer\inst{7},
          Tatiana Podladchikova\inst{8},
          the STIX and EUI teams
          }

   \institute{Institute of Physics, University of Graz, Universitätsplatz 5, 8010 Graz, Austria\\
             \email{stefan.purkhart@uni-graz.at}
             \and
             Kanzelhöhe Observatory for Solar and Environmental Research, University of Graz, Kanzelhöhe 19, 9521 Treffen, Austria
             \and
             Institute for Data Science, University of Applied Sciences and Arts Northwestern Switzerland (FHNW), Bahnhofstrasse 6, 5210 Windisch, Switzerland
             \and
             Institute for Particle Physics and Astrophysics (IPA), Swiss Federal Institute of Technology in Zurich (ETHZ), Wolfgang-Pauli-Strasse 27, 8039 Zurich, Switzerland
             \and
             Space Sciences Laboratory, University of California, 7 Gauss Way, 94720 Berkeley, USA
             \and
             Institute of Physics and Astronomy, University of Potsdam, Potsdam 14476, Germany
             \and
             NorthWest Research Associates, 3380 Mitchell Ln, Boulder, CO 80301, USA
             \and
             Skolkovo Institute of Science and Technology, Bolshoy Boulevard 30, bld. 1, Moscow 121205, Russia
             }

   \date{Received ; accepted }

\abstract
{The Spectrometer Telescope for Imaging X-rays (STIX) onboard Solar Orbiter enables exciting multi-point studies of energy release and transport in solar flares by observing the Sun from many different distances and vantage points out of the Sun-Earth line.}
{We present a case study of an M4-class flare that occurred on March 28, 2022, near Solar Orbiter's first science perihelion (0.33 AU from the Sun). Solar Orbiter had a longitudinal separation of $83.5^\circ$ west of the Sun-Earth line, making the event appear near the eastern limb from its perspective, while Earth-orbiting spacecraft observed it near the disk center.
We follow the evolution of the X-ray and (E)UV sources, analyze their relation to plasma dynamics and heating, and relate our observations to magnetic field structures, including the erupting filament.}
{The timing and location of the STIX X-ray sources were related to the plasma evolution observed in the EUV by the Extreme Ultraviolet Imager (EUI) on Solar Orbiter and the Atmospheric Imaging Assembly (AIA) on the Solar Dynamics Observatory, and to the chromospheric response observed in 1600 Å by AIA. We performed Differential Emission Measure (DEM) analysis to further characterize the flaring plasma at different subvolumes. The pre-flare magnetic field configuration was analyzed using a nonlinear force-free (NLFF) extrapolation.}
{In addition to the two classical hard X-ray (HXR) footpoints at the ends of the flaring loops, later in the event we observe a non-thermal HXR source at one of the anchor points of the erupting filament. These results are supported by a robustness analysis of the STIX images and the co-temporal chromospheric brightenings observed by AIA. 
The full evolution of the AIA 1600~Å footpoints indicates that this change in footpoint location represents a discontinuity in an otherwise continuous westward motion of the footpoints throughout the flare.
The NLFF extrapolation suggests that strongly sheared field lines close to, or possibly even part of, the erupting filament reconnected with a weakly sheared arcade during the first HXR peak. The remainder of these field lines reconnected later in the event, producing the HXR peak at the southern filament footpoint.
Furthermore, we found several thermal X-ray sources during the onset of the impulsive phase and a very low-lying initial thermal loop top source that passes through a double structure during its rise. We were able to relate many of these observations to features of the complex flare geometry involving multiple interacting magnetic flux systems.}
{The combined STIX and AIA observations, complemented by the NLFF extrapolation, allowed us to successfully constrain and verify the signatures of energy release and transport in the flare under study. Our results show that the reconnection between field lines with very different shear in the early phase of the flare plays a crucial role in understanding the motion of the HXR footpoint during later parts of the flare. This generalizes simpler models, such as whipping reconnection, which only consider reconnection propagating along uniformly sheared arcades.}

   
   \titlerunning{Multi-point study of the energy release and transport in the 2022 March 28, M4-flare.}
    \authorrunning{S. Purkhart et al.}

   \maketitle
%

\section{Introduction}\label{sec:introduction}

Solar flares are energy-release events on the Sun that occur over a wide range of energies, often classified as nano-, micro-, and regular-type flares, and can be observed across the whole electromagnetic wavelength range (for observational reviews, see \citealt{Fletcher2011} and \citealt{Benz2017}). The most energetic flares are often accompanied by coronal mass ejections, high-speed particles, and high-energy radiation \citep[e.g., review by][]{Temmer2021}. All of these can profoundly impact the heliosphere, including Earth \citep[e.g., review by][]{Pulkkinen2007}, where they pose a risk to our infrastructure in space and on Earth. Lower-energy flares, while not a direct risk to us, might be an important contributor to the heating of the Sun's multi-million-kelvin corona due to their high occurrence rate \citep{Klimchuk2006,Purkhart2022}, playing a fundamental role in shaping the characteristics of the heliosphere. Therefore, it is crucial to study the flare phenomena in detail to better understand the involved processes and improve our models. Only a detailed understanding of solar flares will allow us to predict their occurrence, impact on the heliosphere, and risks for our life on Earth and in space.

In the current standard flare model, flares are driven by the energy stored in stressed magnetic fields that can be released through a process known as magnetic reconnection \citep{Priest2002,Zweibel2009}. A local reconnection of near anti-parallel magnetic field lines results in the more extensive reconfiguration of the solar magnetic field towards a less stressed state while the released energy is converted into particle acceleration, plasma heating, and bulk motions. The ambient electrons and ions that were accelerated to high energies spiral along the restructured field lines \citep{Miller1997}. Those particles traveling toward lower-lying layers of the solar atmosphere collide with the fields of ambient electrons and ions in the denser chromospheric plasma and lose their energy to heat and bremsstrahlung with typical wavelengths in the hard X-ray (HXR) and gamma-ray range. Observing the HXR spectrum allows us to derive parameters of the accelerated electron population and make conclusions about the acceleration process and plasma conditions \citep{Brown1971,Holman2011}. The rapid heating of the loop footpoints leads to the impulsive evaporation of the chromospheric plasma that fills the closed magnetic loops above, where it gets trapped and cools through conduction and radiation \citep{Neupert1968,Fisher1985,Veronig2005}.
The response of the footpoints to the impulsive heating by decelerated electrons can also be observed through the ultraviolet emission emitted by chromospheric plasma and even white light photospheric emission in the most energetic flares. The emission of hot flare loops that are filled with hot evaporated plasma is mostly visible in the Extreme Ultra Violet (EUV) to soft X-ray energy range and can last up to several hours. Thus, each part of this energy-release and transport sequence in a solar flare emits radiation in a characteristic energy range that we can observe in order to quantify the properties of the different processes involved \citep[e.g., review by][]{Fletcher2011}.

Several instruments observing at different wavelengths have been available to gain more insight into specific elements of the flare phenomena, leading to significant advances in our understanding of the physical processes involved. In particular, multi-instrument flare studies have been of great importance because they allow us to study different aspects of a particular process and also the connection between different processes. Both have strengthened the observational constraints on the standard flare model and helped to refine it. 

With the Solar Orbiter mission \citep{Mueller2020_SolarOrbiter} launched in 2020, we have exciting new additions to the suite of instruments available for multi-instrument flare studies. The Spectrometer/Telescope for Imaging X-rays (STIX, \citealt{Krucker2020_STIX,Benz2012}) onboard Solar Orbiter provides an important continuation for X-ray observations of solar flares and also promises exciting new possibilities through observations much closer to the Sun and from a vast amount of different perspectives compared to previous instruments only observing on the Sun-Earth line. Furthermore, Solar Orbiter carries a suite of four in-situ instruments that enable us to better connect flares with their impact in interplanetary space.

This study focuses on the energy release and transport in the M4-class flare of March 28, 2020. In particular, we demonstrate the potential of STIX X-ray spectroscopy and imaging as part of multi-instrument flare studies \citep[see also the recent studies on microflares during the Solar Orbiter cruise phase by][]{Battaglia2021_FlaresSTIXComissioningPhase,Saqri2022}. The event under study is particularly interesting because Solar Orbiter was near its first perihelion during the nominal science phase, at a distance of only 0.33 AU from the Sun. In addition, Solar Orbiter had a longitudinal separation of $83.5^\circ$ west from the Sun-Earth line. From this perspective, the event was located near the eastern limb of the Sun, while Earth-orbiting instruments observed the event near the disk center.

Our main goal was to investigate different phases of the energy transport using STIX HXR observations and then compare them with the plasma response observed in chromospheric footpoints and with the evolution of the heated plasma in flare loops as observed by the Extreme Ultraviolet Imager (EUI) and the Atmospheric Imaging Assembly (AIA). Since EUI is also onboard Solar Orbiter, it shares the same view of the event as STIX and can add context to the STIX images, while the Earth-orbiting AIA instrument further constrains the observations through a different viewpoint and extensive diagnostic capabilities. Using STIX X-ray observations, we can derive the time evolution of thermal and non-thermal electron populations throughout the flare and determine the origin of thermal and non-thermal X-ray emissions. This enables us to correlate the time and location of the energy input expected from the deceleration of the non-thermal electrons with the chromospheric plasma response in AIA 1600~Å. We track the heated and evaporated plasma further by deriving the emission measure (EM) and temperature evolution of different subregions through differential emission measure (DEM) analysis of AIA EUV images. Finally, we use a nonlinear force-free extrapolation to connect our observations to the magnetic field structures involved in the flare.

\section{Data and Methods}\label{sec:methods}

\subsection{Solar Orbiter/STIX}

STIX \citep{Krucker2020_STIX} is a HXR spectrometer and imager that observes in the 4 to 150 keV energy range. This energy range covers non-thermal bremsstrahlung from the high-energy electrons that are accelerated in the corona during a solar flare and are then rapidly decelerated in the denser parts of the atmosphere by the field of ambient ions. We use STIX observations to derive the location (through imaging algorithms), the spectrum, and the energy content of these non-thermal electrons. In addition, STIX provides diagnostics of the hottest ($>10$ MK) flaring plasma, which we used to study the evolution of the plasma's emission measure and temperature during the flare.

Spectrogram and pixel data for this event were downloaded directly from the STIX website\footnote{\url{https://datacenter.stix.i4ds.net}}.
We corrected all data for the difference in light travel time from the Sun of 335.7 s between the Solar Orbiter spacecraft and an Earth-based observer. Therefore, all times in this paper (STIX light curves, spectrograms, and images) are given for an Earth-based observer, making it easier to compare the data with the AIA observations.

The measured X-ray count spectrum was fitted with the expected emissions from different electron populations using the OSPEX\footnote{\url{http://hesperia.gsfc.nasa.gov/ssw/packages/spex/doc/}} software, which is part of the SolarSoftWare environment for the IDL programming language (SSWIDL). A pre-event background was subtracted to isolate the count enhancements due to the observed event. We assumed that the remaining spectrum was the result of either a purely thermal electron distribution or a combination of thermal and non-thermal electron distributions. We fitted the thermal part of the spectrum with the isothermal model 'vth' and the non-thermal energy part with the thick target model 'thick2', which we restricted to a single power law. In cases where we fitted only the isothermal model because no satisfactory non-thermal fit could be obtained, we restricted the upper end of the fitted energy range to 12 -- 15 keV to obtain a consistent EM and temperature evolution.  

This spectral fitting procedure was performed for a sequence of 20 s intervals spanning the impulsive phase of the flare. From this sequence of spectral fits, we derived the time evolution of essential parameters of the thermal and non-thermal electron populations during the flare. Furthermore, we performed additional spectral fitting for individually selected intervals focusing on times of impulsive energy release, apparent by impulsive increases of STIX counts in the 25 -- 50 keV and 50 -- 84 keV light curves.

STIX imaging was performed for those selected intervals using standard imaging algorithms available in the SSWIDL distribution, including the Clean method \citep[stx\_vis\_clean,][]{Hogbom1974}, the maximum entropy method (MEM\_GE) \citep[stx\_mem\_ge,][]{Massa2020}, and forward fitting \citep[stx\_vis\_fwdfit\_pso,][]{Volpara2022}. A constant shift of $\Delta x=-50$ arcsec and $\Delta y=15$ arcsec in helioprojective coordinates had to be applied to all STIX images to align non-thermal footpoint contours with AIA 1600 contours reprojected to the Solar Orbiter perspective. This constant shift was needed because the aspect solution included within the SSWIDL package provided insufficient correction for this particular event. Derived STIX non-thermal imaging contours were reprojected to the AIA perspective for comparison with EUV observations. Since the STIX thermal X-ray sources usually lie within the higher atmosphere (flare loops), we did not reproject them. Instead, we constructed a line-of-sight from Solar Orbiter to each local maximum within the 70\% thermal contours. This line-of-sight was then reprojected to the AIA perspective instead of the contours. All reprojections and visualizations utilized version 4.0.4 \citep{SunPy_4.0.4} of the SunPy open source software package \citep{sunpy_community2020} for the Python programming language.

\subsection{Solar Orbiter/EUI}

For EUV observations that match the STIX perspective, we used data from the Extreme Ultraviolet Imager \citep[EUI,][]{Rochus2020_EUI} onboard Solar Orbiter. The instrument consists of three telescopes: the Full Sun Imager (FSI) and two High Resolution Imagers (HRI). HRI captures high resolution (two pixels 1 arcsec), high cadence (as low as 1s for the full FOV) images of the upper chromosphere to low corona (Lyman $\alpha$ and 17.4 nm), but its narrow FOV did not cover the flare location. We, therefore, used images from the FSI to directly compare the STIX observations with the plasma evolution seen from the same perspective. FSI produces $3.8^\circ \times 3.8^\circ$ images (a width of four solar radii at perihelion) in 174 and 304~Å pass bands with an angular resolution of 10 arcsecs, corresponding to a linear resolution of about 2.4 Mm on the Sun at the time of the event. Since the cadence was limited to 10 min due to telemetry constraints and the fixed exposure time of typically a few seconds leads to overexposure during the flare, the FSI images were mainly used as reference images for the STIX imaging contours and not for detailed analysis of the flaring plasma.

EUI data was made available through the Solar Orbiter EUI Data Release 5.0 2022-04 \citep{EUI_datarelease5.0} and was downloaded from the EUI website\footnote{\url{https://www.sidc.be/EUI/data/}} as L2 data. This processing level already includes corrections for some instrument deviations and updated WCS FITS keywords to correct spacecraft pointing instabilities. As with the STIX data, all times were corrected for the light travel time between Solar Orbiter and Earth for easier comparison with other observations from near Earth.

\subsection{SDO/AIA and HMI}

For a detailed analysis of EUV and UV observations, we used data from AIA \citep{Lemen2012_AIA} onboard the Solar Dynamics Observatory \citep[SDO,][]{Pesnell2012_SDO} that has orbited the Earth in a geosynchronous orbit since 2010. The instrument captures full disc images of the Sun in multiple wavelengths with a 1.5 arcsec spatial resolution at a pixel scale of 0.6 arcsecs and a standard operating cadence of 12 s.

We used the six coronal EUV wavelength channels (94, 131, 171, 193, 211, and 335~Å) corresponding to iron lines at various ionization stages (Fe\,{\sc xviii}, Fe\,{\sc viii} and Fe\,{\sc xxi}, Fe\,{\sc ix}, Fe\,{\sc xii} and Fe\,{\sc xxiv}, Fe\,{\sc xiv} and Fe\,{\sc xvi}). The wavelength channels are not only sensitive to the peak formation temperatures of their lines (log T[K]: 6.8, 5.6 and 7.0, 5.8, 6.2 and 7.3, 6.3 and 6.4). Rather, they are all characterized by a much broader temperature response, making the chosen selection of EUV wavelengths sensitive to plasma temperatures in the range of at least $10^5$ to $10^7$ K \citep{Lemen2012_AIA,Boerner2012}. Combined with differential emission measure analysis, we can obtain accurate temperature and emission measure estimates for the flaring plasma from the AIA EUV images.

In addition, we used images from AIA's 1600~Å UV filter, which is sensitive to chromospheric plasma temperatures of about $10^5$ K. Image sequences from this filter were used to analyze the chromospheric plasma response at the flare footpoints, where the energy from high-energy electrons is deposited. We extracted the time evolution of total count rates within specified sub-regions and subtracted a pre-event background to isolate count enhancements from the flare.
AIA 1600~Å images were also reprojected to the Solar Orbiter perspective to confirm co-alignment with STIX non-thermal X-ray contours.
Finally, we also used images from the AIA 304~Å channel for analyzing STIX imaging contours because of their similarity with the EUI 304~Å filter.

AIA full disk images were downloaded from JSOC\footnote{\url{http://jsoc.stanford.edu}} as level 1 data. We deconvolved all EUV images using the $aia\_deconvolve\_richardsonlucy$ SSWIDL function with default settings and the point spread function obtained for all wavelengths through the $aia\_calc\_psf$ function. This procedure reduces the stray light from localized bright areas present during a solar flare, drastically improving contrast and sharpness in the reconstructed DEM maps and enabling us to get better plasma diagnostics for different submaps. Further processing of the deconvolved data to level 1.5 and the creation of submaps was performed using standard SSWIDL procedures. All further visualizations of AIA images and reprojections to the Solar Orbiter perspective utilized SunPy.

Line-of-sight magnetograms from the Helioseismic and Magnetic Imager \citep[HMI;][]{Schou2012_HMI} were used to relate flare ribbons and footpoints observed by AIA and STIX, respectively, to the underlying magnetic field. In addition, we used vector magnetograms for the nonlinear force-free magnetic field extrapolation (see section \ref{sec:methods_NLFF}).

\subsection{Differential emission measure analysis}

Since the chosen AIA EUV wavelengths are sensitive to the emission from the solar corona, an optically thin plasma that can be assumed to be in thermal equilibrium, the intensities $I_\lambda$ in each wavelength channel $\lambda$ can be described as

\begin{equation}
    I_\lambda = \int_T K_\lambda(T) DEM(T) dT,
\end{equation}\\
with the filters response function $K_\lambda$ and the differential emission measure $DEM$, both given as a function of the plasma temperature $T$. The DEM is defined as a function of the electron density $n$ along the line-of-sight $h$ as

\begin{equation}
    DEM(T) = n(h(T))^2dh/dT
\end{equation}\\
and is, therefore, a measure of the temperature distribution of radiating plasma along this line of sight. The desired DEM cannot be calculated directly from the measured counts in each pixel since it is convolved by the instrument response and the emission process. Consequently, multiple algorithms with different approaches have been developed to reconstruct the DEM from observations.

To derive the DEM from the chosen AIA EUV observations, we used the IDL implementation of the regularized inversion algorithm developed by \citet{Hannah2012}. We prepared map structures containing images from the six selected AIA EUV wavelength channels (94, 131, 171, 193, 211, and 335~Å) that were binned by 2x2 pixels. The advantage is a better signal-to-noise ratio and, consequently, more accurate and more stable DEM results. This binning of pixels resulted in an effective spatial resolution of 1.2 arcsecs per pixel for the DEM results. Additionally, all AIA images used for DEM reconstruction were differentially rotated to 10:50:04~UT to improve alignment. The temperature range for DEM reconstruction was set from 0.5 to 31.6 MK and divided into 40 evenly spaced logarithmic temperature bins. We calculated the AIA temperature response function for the observation time using the \textit{aia\_get\_response} SSWIDL routine using coronal abundances from the CHIANTI 10.0 database  \citep{Dere1997_CHIANTI_1,Del_Zanna2021_CHIANTI_10}.

During the observations, AIA gradually switched some EUV channels into flare mode, which takes image series that alternate between the regular and a shorter exposure time. The shorter exposure time reduces overexposure and blooming effects during the flare. Consequently, we used only the image series with the lower exposure time for DEM reconstruction to improve the results. Unfortunately, this doubles the effective time cadence of the DEM reconstructions to 24 s.

A background DEM was subtracted from all other DEM reconstructions to isolate the distribution of the flaring plasma. From the background-subtracted DEM, we calculated for selected subregions in the flare the total emission measure ($EM$) of all temperature bins ($\Delta T$) by

\begin{equation}
  EM = \sum DEM(T) \cdot \Delta T
\end{equation}\\
and the emission weighted temperature ($\bar{T}$) \citep[cf.][]{Cheng2012,Vanninathan2015} as

\begin{equation}
  \bar{T}=\frac{\sum(DEM(T) \cdot \Delta T \cdot T)}{\sum(DEM(T) \cdot \Delta T)}
\end{equation}

\subsection{Nonlinear force-free magnetic field extrapolation}
\label{sec:methods_NLFF}

A nonlinear force-free magnetic field extrapolation was performed using the approach developed by \citet{Jarolim2023}.
A physics-informed neural network uses the force-free assumption and observations of the photospheric vector magnetic field to derive a solution for the coronal magnetic field.
Using this method, we performed the extrapolation for the cylindrical equal area (CEA) projection of the entire Spaceweather HMI Active Region Patch (SHARP) 8088 at 11:00~UT, which includes the AR 12975 under study and AR 12976 further east.
The obtained extrapolation shows the pre-event magnetic field configuration, with the aim of better understanding which magnetic field structures may have gotten involved in the eruption/flaring process.
The full code and an interactive Google Colab notebook to run the magnetic field extrapolations for any active region are available on the project's Github\footnote{\url{https://github.com/RobertJaro/NF2}}.

To visualize the results, we imported the extrapolation into the ParaView\footnote{\url{https://www.paraview.org}} open-source visualization software. We overlayed the 11:00~UT HMI line-of-sight (LOS) magnetogram with the contours of the AIA 1600 Å running difference images, taking into account the differential rotation during the time difference between the magnetogram and the HXR peak. The resulting images for each HXR peak were used as the base layer for visualization in ParaView. Using this setup, we placed seed sources for stream tracers inside the AIA 1600~Å contours to draw the magnetic field lines connecting to the flare kernels of interest.

\subsection{GOES}

We used data from the Geostationary Operational Environmental Satellites (GOES) observed in the 0.5 -- 4~Å and 1 -- 8~Å SXR bands. We subtracted a pre-event background and calculated the EM and temperature evolution using the standard routines available in the SSWIDL distribution \citep{White2005}, which were updated in June 2022.

\section{Results}\label{sec:results}

\subsection{AIA and HMI event overview}\label{sec:results_aia}

\begin{figure*}
\centering
  \includegraphics[width=18cm]{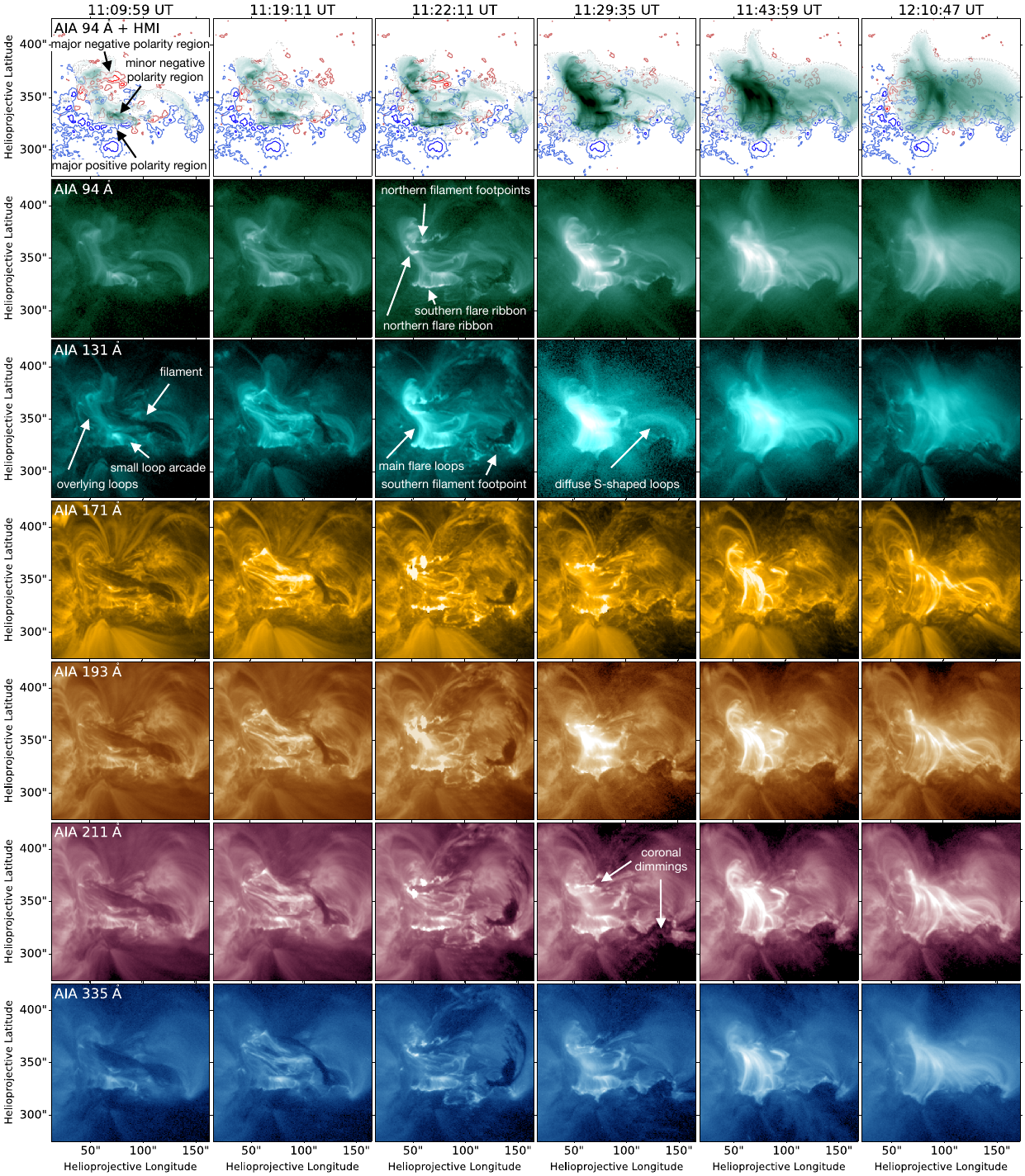}
    \caption{
    Event overview in six AIA EUV wavelength channels at six selected times (columns). Images are normalized by exposure time and scaled to a fixed minimum and maximum value for each channel. The first row shows inverted AIA 94~Å images together with HMI line-of-sight magnetogram contours marking levels of $\pm 200$ and $\pm 1000$ G for negative (red) and positive (blue) polarity. Movie: \href{https://drive.google.com/drive/folders/1ZiBDdO1Lk6x_q6X_3Y7LFT8V7NyQQQ3H?usp=sharing}{Google Drive}
    }
    \label{fig:AIAoverview}
\end{figure*}

Figure \ref{fig:AIAoverview} shows an overview of the event as captured by the six coronal AIA EUV wavelength channels. The included frames were selected to highlight different phases of the event, including the filament before and during the eruption, the peak of EUV emission at the top of the loops, and the further evolution of the flare loop arcade. A video accompanying this figure shows the event at the full 12-second AIA cadence, allowing viewers to follow all phases of the event in greater detail. In addition, the figure includes line-of-sight magnetograms from HMI for the selected time steps, overlaid with inverted AIA 94~Å to illustrate the magnetic polarities involved in each substructure. 

Active region 12975 was classified as a $\beta$ magnetic field configuration by the NOAA/USAF Active Region Summary issued on March 28, 2020, 00:30~UT, and as a D-type sunspot configuration by the Kanzelhöhe Observatory for Solar and Environmental Research \citep[KSO,][]{Poetzi2021_KSO}\footnote{\url{http://cesar.kso.ac.at/main/cesar\_start.php?date=2022 -- 03-28}}.

The images taken around 11:09:59~UT, during the pre-eruptive phase of the event, show the filament as a dark, elongated structure against the brighter plasma of the active region. As the filament slowly rises, a small loop arcade below shows several bursts of brightening at different locations, which may be related to the slow rise and destabilization of the filament until it is impulsively accelerated and ejected. This small loop arcade, formed in a confined C-class flare at 09:55~UT, connects parts of the major positive polarity region with the adjacent parallel minor negative polarity region. The filament connects from another region of positive polarity in the southwestern portion of the AR toward the major negative polarity region. Some overlying loops can be seen in the 131~Å channel that connect the major negative polarity region with parallel parts of the positive polarity region just south of the small flare loop arcade.

Note that this filament configuration only existed within the last hour before the eruption/flare under study. Its original configuration changed during a C-class flare at 09:55~UT. Thus, the shown filament configuration was not particularly stable and soon erupted after its reconfiguration. However, the formation and destabilization of this filament are beyond the scope of this paper and will be the subject of a separate study.

Around 11:19:11~UT, the impulsive acceleration and eruption of the filament can be observed. At the same time, the small loop arcade and its ribbons further increase in brightness, and additional large flare loops appear in channels sensitive to the hottest plasma (e.g., 94 and 131~Å). These connect the major negative polarity region with the positive flux immediately south of the small flare loop arcade. Sections of the erupting filament plasma also brighten during this phase, visible in all AIA channels, and there is an increase in the emissions from the southern filament footpoint. The filament eruption appears asymmetric, with a more active leg in the north and a more anchored leg in the south (see Fig. \ref{fig:AIAoverview} and the accompanying movie). This asymmetric behavior can be also seen in the side-view by EUI. We note that the filament eruption is associated with a halo CME with a speed of approximately 700 km s$^{-1}$ as reported by the SOHO/LASCO CME catalog\footnote{\url{https://cdaw.gsfc.nasa.gov/CME_list/}} \citep{Yashiro2004}.

As the erupting filament continues to be ejected from the active region, it can be seen as an arc-like structure (e.g. 11:22:11~UT). We observe multiple brightenings in all AIA channels, which seem to correspond to the position of the flare ribbons and kernels. The southern filament footpoint area appears to follow the semicircular shape of the positive polarity region, while the northern filament footpoints appear to be distributed across several kernels of the EUV emission in the major negative polarity region. The northern flare ribbon extends from the brightest emission kernel in the negative polarity region. Flare loops connect this kernel to the southern flare ribbon in the positive polarity region, which forms immediately southward of the positive ribbon of the small loop arcade. Both progress jointly southward. In addition, we see bright loops extending from the flaring region over the northern flare footpoint and connecting towards the weaker northern kernels.

The initially very narrow flare loop arcade expands westward. Simultaneously its brightness increases and the loop top emission reaches a maximum in the 131~Å channel around 11:29:35~UT. The AIA channels sensitive to cooler plasma also increasingly show flare loop emissions. In addition, a set of diffuse S-shaped hot loops connecting from the flaring region towards the southern filament footpoint brightens and we see coronal dimmings \citep[best visible in the 171, 193, and 211~Å channels; e.g.][]{Dissauer2018} extending north and southwest of the active region, in the areas of the northern and southern filament footpoints. These are partly surrounded by the S-shaped end sections of the flare ribbons.

As the flare progresses (e.g., 11:43:59~UT and 12:10:47~UT), the large flare loop arcade grows as the flare ribbons move apart and away from the polarity inversion line (PIL). A continued westward growth of the flare arcade, extending up to the southern filament footpoint area in the 131 and 94 A channels, is also observed during this period.

\subsection{STIX hard X-ray imaging and spectroscopy}\label{sec:results_stix}

\begin{figure*}
\centering
  \includegraphics[width=17.77cm]{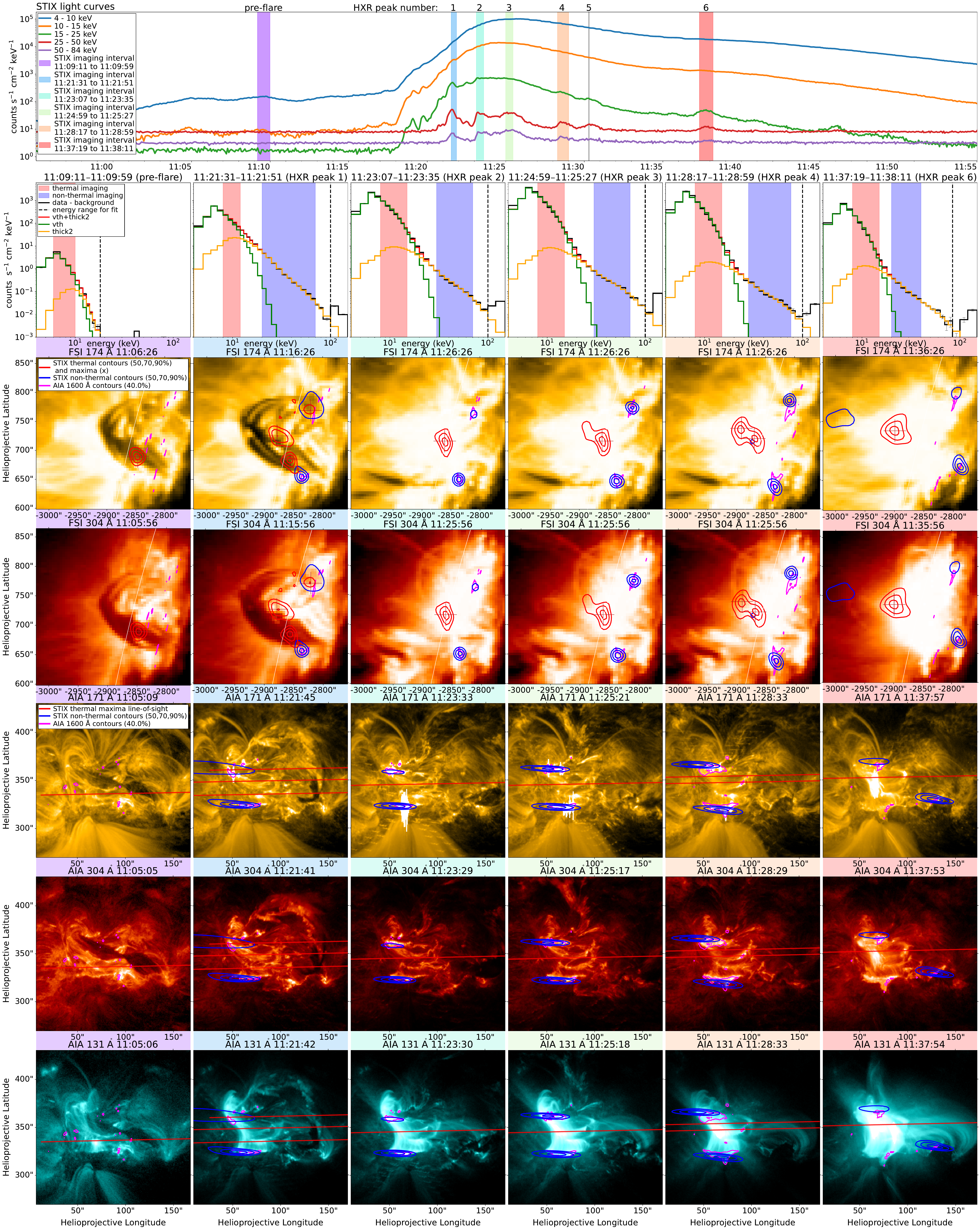}
    \caption{
    STIX observations of the event and comparison with EUI and AIA images.
    Top row: STIX light curves for five energy ranges with marked time intervals for selected HXR peaks. 
    Columns: Color-coded corresponding to the marked time intervals containing the following observations: 
    1) STIX X-ray spectrum fitted with a thermal (vth) and non-thermal (thick2) electron model. Energy intervals used for imaging are marked. 
    2) STIX clean image contours on top of the closest available EUI FSI 174~Å image. Maxima of thermal contours are marked. 
    3) Same as 2) but with EUI FSI 304~Å image. 
    4) AIA 171~Å image with reprojected STIX non-thermal contours and the line-of-sights through the maxima of STIX thermal sources. 
    5) and 6) as 4) but with AIA 304 and 131~Å images, respectively.
    }
    \label{fig:STIXoverview}
\end{figure*}

In Figure \ref{fig:STIXoverview}, we present an overview of our STIX imaging and spectral analysis for six selected time intervals. The first interval covers a thermal peak during the pre-flare phase, while the other five contain major HXR bursts (labeled as HXR peaks 1, 2, 3, 4, and 6) that indicate times of significant energy release and particle acceleration. We show the measured count spectra for each interval together with the modeled count spectra from the fitted thermal and non-thermal electron populations. The detailed time evolution of all fit parameters of the thermal and non-thermal electron populations can be seen in Fig. \ref{fig:DEMtimeseries}, which is included in a later section of this paper. Reconstructed Clean images of the STIX observations and comparison with both EUI and AIA images show the evolution of two non-thermal HXR sources at the footpoints of the flare and thermal emission from the flare loops. The STIX images for the last HXR peak (HXR peak 6) show a change in the non-thermal footpoint configuration, with the main non-thermal X-ray source located at the southern filament footpoint. In this section, we will discuss each of the HXR peaks individually and highlight additional interesting observations.

During the pre-flare interval (11:09:11 to 11:09:59~UT), the STIX images derived from the 6 -- 10 keV energy range reveal a single thermal source whose line-of-sight matches the small loop arcade seen by AIA 131~Å. Our fit to the observed X-ray spectrum clearly shows that the source is dominated by thermal emission ($EM = 0.02 \times 10^{49}$ cm$^{-3}$ and $T = 10.1$ MK). The shown spectral fit includes a steep non-thermal component (electron spectral index $\delta>10$) at higher energies, but the spectrum could also be well-fitted with only a thermal component. Those higher energies were, however, unsuitable for reliable image reconstruction. Therefore, we only include the thermal contours for the 6 -- 10 keV energy range in this time interval. The EUI images shown were captured about three minutes before the STIX and AIA observations. Due to the strong projection effects close to the limb, the filament appears almost loop-like in the EUI images. However, the AIA images reveal its real elongated shape.

The observed STIX spectrum during HXR peak 1 (11:21:31 to 11:21:51~UT), which corresponds to the first major HXR peak, shows clear non-thermal emission with a spectral index of $\delta=4.8$ dominating the X-ray spectrum from about 20 to 100 keV. The thermal component has risen significantly in both EM ($0.08 \times 10^{49}$ cm$^{-3}$) and temperature (19.0 MK). For imaging, we selected the 8 -- 12 keV energy range from the thermal part of the spectrum and the 20 -- 70 keV range of the non-thermal spectrum. The STIX imaging contours show two non-thermal footpoints, with the southern one being much stronger and more compact than the northern more diffuse footpoint. Multiple thermal sources lie in between the two non-thermal footpoints. Comparison with the corresponding EUI images suggests that the locations of the non-thermal sources are close to the footpoints of the erupting filament, while all thermal sources lie somewhere beneath the filament. It is important to note that the EUI images included in this interval show the filament's position about five minutes before the STIX and AIA observations due to their limited cadence of 10 minutes. The filament in these EUI images has clearly risen in height compared to the previous image and still appears loop-like.
Reprojected STIX non-thermal contours enable us to better match the non-thermal sources to various flare features. We find that the southern non-thermal source closely matches the location of the southern flare ribbon indicated by the AIA 1600~Å contours and by the flare loops seen in AIA 131~Å. The reprojected northern non-thermal source covers the general area of the northern flare and filament footpoints, but cannot be assigned to a single feature. The southernmost thermal line of sight again aligns with the smaller loop arcade that was already active before the eruption. The middle thermal source crosses the top of the large developing flare arcade, while the northern thermal source lies just above the northern flare and filament footpoints with their diffuse non-thermal source. The AIA images reveal the true state of the filament eruption during the investigated HXR peak.

The observed non-thermal emission hardens during HXR peak 2 (11:23:07 to 11:23:35~UT), reaching a spectral index of $\delta=4.2$, while the thermal EM and temperature rise further to $EM = 0.20 \times 10^{49}$ cm$^{-3}$ and $T = 21.6$ MK. In the derived STIX non-thermal images (30 -- 70 keV), we observe a more localized northern footpoint that now closely matches the AIA 1600~Å contours of the northern flare ribbon. We will thus refer to this non-thermal source as the northern flare footpoint from now on, but readers should keep in mind that the location is also very close to the northern footpoint of the erupting filament. The southern non-thermal flare footpoint stays mostly the same as during HXR peak 1. The EUI images shown are taken three minutes after HXR peak 2. We still find a large and a small flare arcade in the AIA 131~Å images, but the loop top source of the large arcade clearly outshines the smaller arcade in STIX thermal images (8 -- 15 keV). Interestingly, the thermal loop top source lies significantly lower than the loop top position indicated during the HXR peak 1. The relative positions of the two non-thermal flare footpoints and thermal loop top source, as seen from the Solar Orbiter perspective, suggest quite a flat loop profile.

During HXR peak 3 (11:24:59 to 11:25:27~UT), the non-thermal emission hardens further and reaches a spectral index of $\delta=3.8$. The EM of thermal electrons continues rising to $0.43 \times 10^{49}$ cm$^{-3}$ while their temperature starts to decrease again to $19.6$ MK. We observe only minimal change in the position of the thermal maximum (8 -- 15 keV) that remains at low heights. However, the reconstructed STIX images suggest the possible extension toward a second thermal source higher up. The EUI images shown are taken one minute after the HXR peak. We find that the thermal maxima's line-of-sight still intersects the brightest region at the loop top in the AIA 131~Å images. The northern non-thermal source (30 -- 70 keV) is now fully developed, matching the strength and compactness of the southern flare footpoint. Both reprojected non-thermal contours closely match the AIA observations.

HXR peak 4 (11:28:17 to 11:28:59~UT) appears a bit separated from the main group and occurred about 3 minutes after HXR peak 3. The fitted non-thermal spectrum hardens only slightly to a spectral index of $\delta=3.7$ while thermal EM still rises ($0.51 \times 10^{49}$ cm$^{-3}$) and temperature further decreases (16.9 MK). The thermal source (8 -- 15 keV) has evolved into a clear double structure with two separate maxima. In addition, we find some indications of non-thermal emission (30 -- 70 keV) from the same region that is partially visible within the 50\% contours shown here. However, those sources can not be well constrained due to a low count number during this interval and could be attributed to noise. These non-thermal loop sources start to fade when we increase the lower energy bound to 40 keV for image reconstruction.
The higher (more northern from the AIA perspective) thermal line-of-sight has moved away from the brightest loop region, as seen in 131~Å. We find that the non-thermal sources from both flare footpoints are still compact and have moved apart, away from the PIL. 
The northern flare footpoint also shifts slightly away from the AIA-observed loop footpoints towards the East. However, one has to keep in mind the very flat projection angle (about $70^{\circ}$) that amplifies shifts in the east-west direction but does not significantly affect STIX's accuracy in the north-south direction when reprojecting sources to the AIA perspective. The shown EUI images now match the time of the STIX and AIA observations. The filament has been ejected and has left the shown FOV, but its connection to the filament footpoints remains visible.

The last HXR peak (HXR peak 6; 11:37:19 to 11:38:11~UT) shows a slightly softer non-thermal spectrum (electron spectral index $\delta=4.0$) and a significantly reduced EM ($0.30 \times 10^{49}$ cm$^{-3}$) and temperature (14.5 MK). STIX imaging reveals a single thermal source (8 -- 15 keV) that has risen in height compared to both parts of the double structure in the previous time interval. Comparison with AIA shows that the thermal line-of-sight now matches the brightest region of the loop top again. STIX non-thermal images (20 -- 40 keV) show an interesting change in the source configuration. We located the major non-thermal source at the southern anchor point of the erupted filament. This location lies at the edge of a coronal dimming that is thought to mark the footpoint of the erupted flux rope \citep[e.g.][]{Sterling1997,Veronig2019}. In addition, we find weaker emission from the northern flare footpoint, which coincides with the northern filament footpoint, as well as a coronal source.

\begin{figure}
  \resizebox{\hsize}{!}{\includegraphics{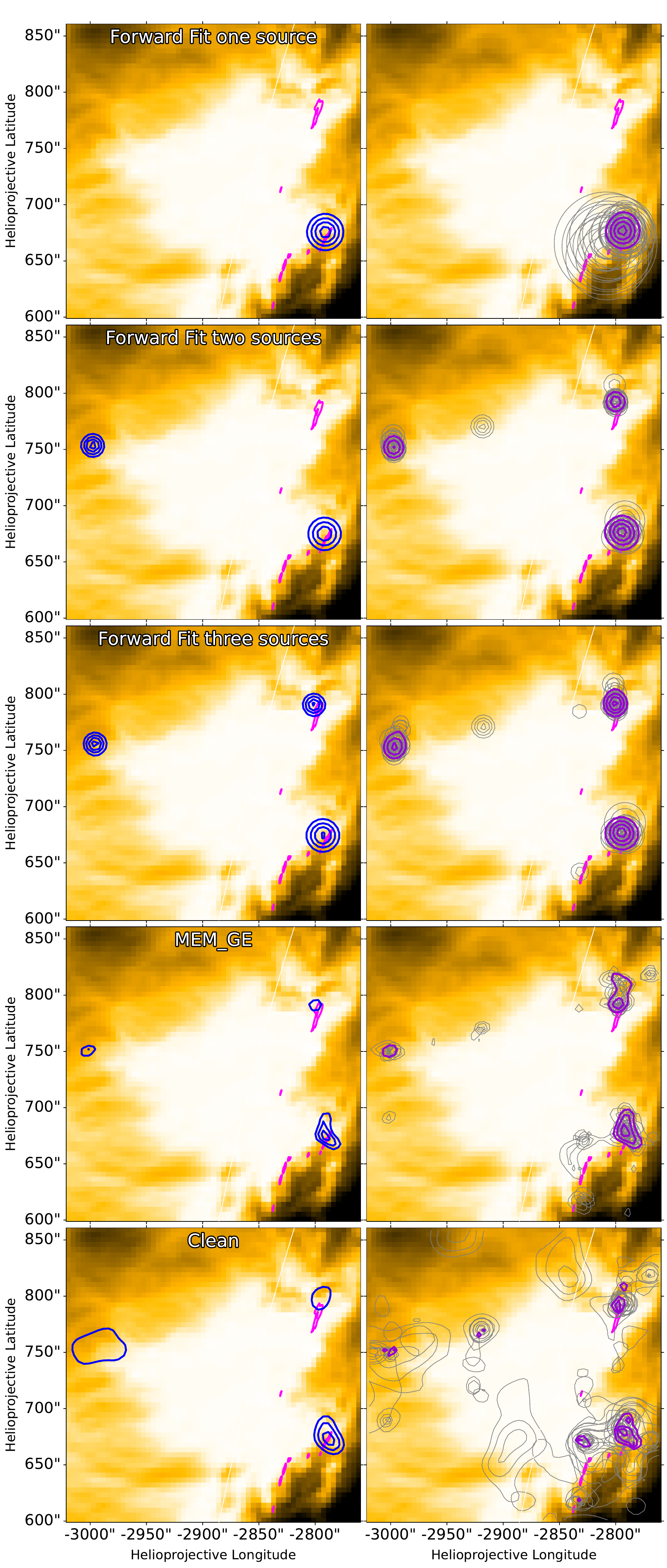}}
  \caption{
  Robustness analysis for STIX images (20 -- 40 keV) of the late HXR peak (HXR peak 6; 11:37:19 to 11:38:11~UT).
  Left, from top to bottom: Unperturbed image contours (blue) from forward fitting a model of one, two, and three circular sources, respectively, MEM\_GE, and Clean on top of the closest EUI FSI 174~Å image (same as Fig. \ref{fig:STIXoverview}) and AIA 1600~Å contours.
  Right: Image contours of 20 reconstructions from perturbed visibilities (gray) and their mean (purple). Forward fit contours show 30, 50, 70, and 90\% levels, while MEM\_GE and Clean show 50, 70, and 90\% levels.
  }
  \label{fig:perturbations_overview}
\end{figure}

STIX measured 5158 counts above 8389 background counts in the selected interval (11:37:19 to 11:38:11~UT) and energy range (20 -- 40 keV), giving a ratio of only about 1.6 between total and background counts. Thus it is possible for noise to significantly affect the reconstructed images. To check the robustness of the STIX imaging during HXR peak 6, we followed a procedure recently applied by \citet{Stiefel2023} based on the confidence strip method \citep{Piana2022}. We added Gaussian noise with a standard deviation equal to the error on the visibility amplitudes to the observed visibilities before applying the Forward Fit, MEM\_GE, and Clean imaging methods. This procedure was repeated for 20 different random noise distributions. Fig. \ref{fig:perturbations_overview} shows the contours of all perturbed images and their mean for the different imaging methods and compares them to the original unperturbed images.
 
Forward fit was used to fit models consisting of one, two, or three circular sources, respectively. For each source, we set the lower limit of the full width at half maximum (FWHM) to 15 arcsec, which is the nominal resolution of the finest STIX subcollimator used. Allowing even smaller sources did not change the general result of this robustness analysis, but the mean contours were sometimes dominated by a few very compact sources with high local counts. We find that in all 20 perturbed images, the single source is placed around the southern filament footpoint and is often rather large, with its center varying around the unperturbed source position. The mean contours agree well with the source reconstructed from the unperturbed visibilities. 
Choosing a model with two circular sources, one source always remains at the southern filament footpoint, whereas the second source is 10 times placed on the coronal and nine times at the northern flare footpoint (including the one source placed slightly too far north of the actual flare footpoint location). In one of the 20 reconstructions, the second source is placed at a different location between the northern flare footpoint and the coronal source.
If we allow a third source in the model, the southern filament footpoint is found in each fit, the coronal source 19 times, and the northern flare footpoint 18 times. The approximate location of all three sources (within about 25 arcsecs of the original) was simultaneously found in 17 perturbations. Errors include the same source between the northern flare footpoint and the coronal source that was also found in the reconstructions with just two fitted sources. Another source on the same line but located closer to the northern flare footpoint is found once. Another time a source is placed near the southern flare footpoint.

The mean contours of the MEM\_GE reconstructions include all three sources within their 50\% level, but the location of the northern flare footpoint becomes much more extended compared to the original. In individual perturbed images, the size of all sources is often significantly altered, and their location is shifted. The original southern filament footpoint location is included within 50\% contours of the perturbations in 19 out of 20 cases, the coronal source only in about 6, and the northern flare footpoint in about 13 reconstructed images.
For the Clean images shown, the residuals are added to the final images after reconstruction, which results in increased noise compared with the other methods. 50\% levels of the mean contours still include the original sources but also include other regions near the footpoints and the coronal source already discussed for the forward fit method.

In summary, the southern filament footpoint location is the most consistent throughout the applied perturbations and is present in all forward fit reconstructions. While it is also included in almost all MEM\_GE and Clean reconstructions, it is often strongly distorted and shifted, indicating that the source is only slightly above the noise level.
The other sources are less reliable. The coronal source was not found by the forward fit in two and the northern flare footpoint in three of the 20 perturbations modeled with three sources. In MEM\_GE and Clean reconstructions, the northern flare footpoint appears more consistent than the coronal source.

We conclude that the reconstructed images are at the limit of STIX's imaging capabilities. However, the major source of the detected non-thermal X-rays can be constrained to the vicinity of the southern filament footpoint. The two other sources are close to the noise level and consequently more uncertain.

\subsection{Chromospheric response in UV}\label{sec:results_1600}

\begin{figure*}
\centering
  \includegraphics[width=18cm]{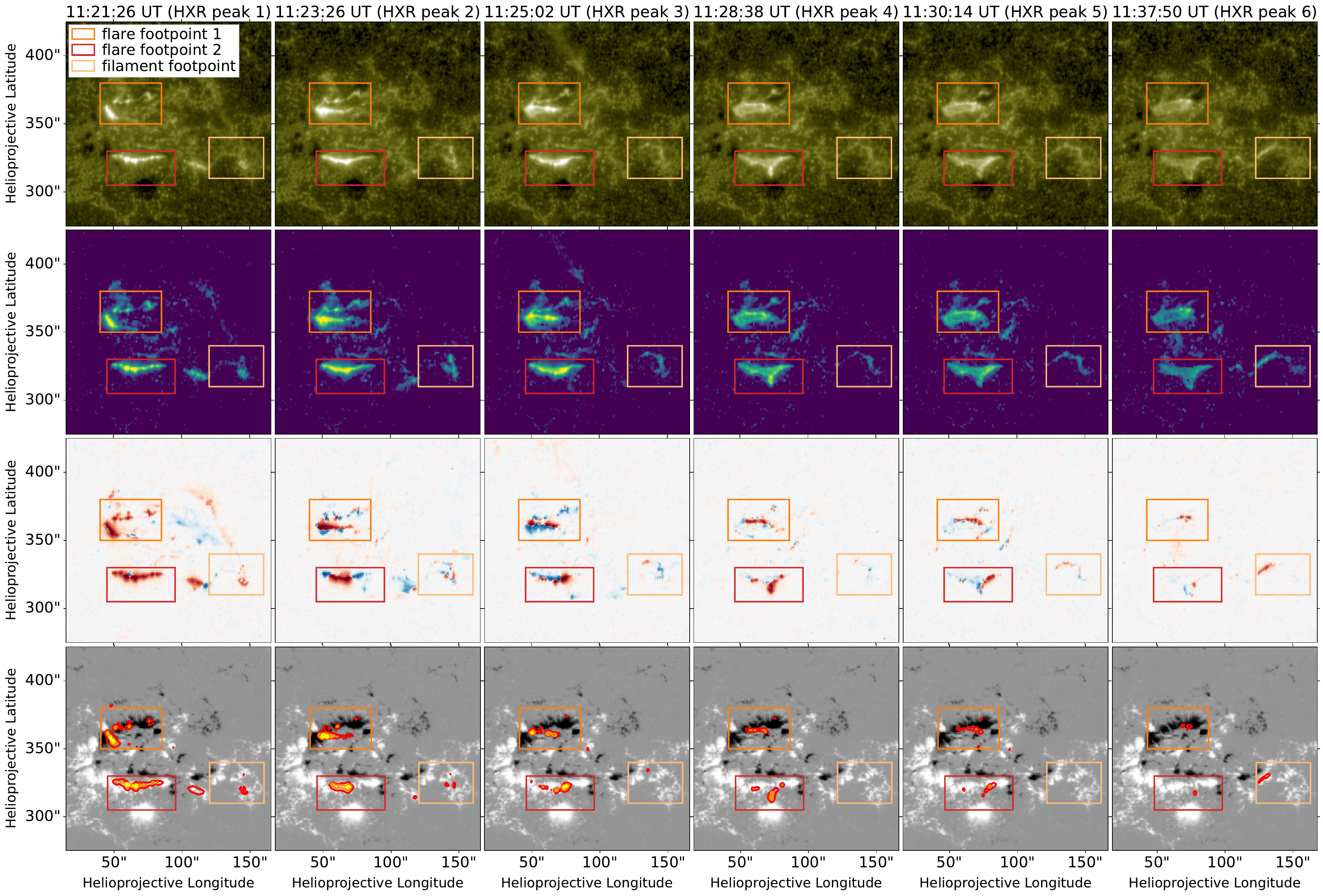}
    \caption{
    AIA 1600~Å observations and their relation to the HMI LOS magnetic field.
    Top row: AIA 1600~Å images taken during HXR peaks detected by STIX. The observation times and corresponding HXR peak numbers are given on top. Colored rectangles mark subregions encompassing the northern footpoints of the filament and northern flare ribbon (orange), the southern flare ribbon (red), and the southern filament footpoint (yellow).
    Second row: Base difference images of the frames shown on top relative to the AIA 1600~Å image taken at 11:00~UT.
    Third row: Running-difference images of the above frames relative to the AIA 1600~Å image taken 48 s earlier. Red and Blue indicate positive and negative change, respectively.
    Bottom row: Contours of positive changes (emission increase) in the running difference images above on top of the HMI line-of-sight magnetograms.
    Movie: \href{https://drive.google.com/drive/folders/1ZiBDdO1Lk6x_q6X_3Y7LFT8V7NyQQQ3H?usp=sharing}{Google Drive}}
    \label{fig:1600images}
\end{figure*}

\begin{figure}
  \resizebox{\hsize}{!}{\includegraphics{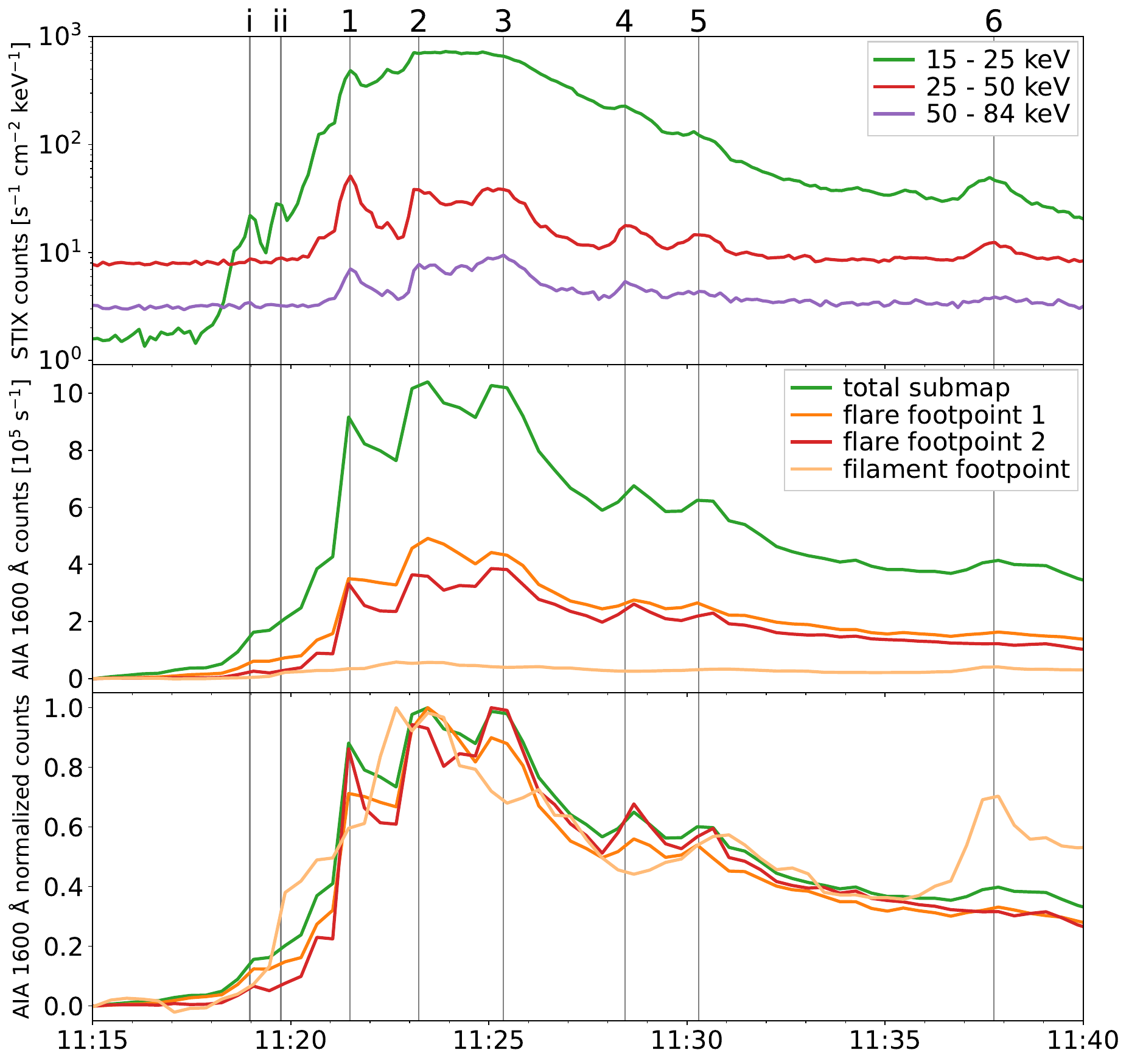}}
  \caption{
  Time correlation between STIX HXR counts and AIA 1600~Å counts in selected subregions.
  Top panel: STIX count flux in three different energy ranges. Times of prominent HXR peaks are marked by vertical lines and labeled on top. 
  Middle panel: Total AIA 1600~Å counts for subregions marked in Fig. \ref{fig:1600images}. 
  Bottom panel: Normalized total AIA 1600~Å counts for the same subregions.}
  \label{fig:1600timeseries}
\end{figure} 

As a further check, we examined the chromospheric UV emission in AIA 1600~Å. In Fig. \ref{fig:1600images} we present a selection of AIA 1600~Å images during the main HXR bursts detected by STIX. They largely correspond to the peaks previously analyzed in section \ref{sec:results_stix}, except that we exclude the pre-impulsive thermal peak (11:04:39 to 11:05:35~UT in Fig. \ref{fig:STIXoverview}). Instead, we include the non-thermal peak at about 11:30:14~UT (HXR peak 5). This image series shows the chromospheric response to the energy deposited by the non-thermal electrons in the lower atmosphere. Fig. \ref{fig:1600images} also includes base difference images to further highlight regions of increased emission since the onset of the flare, and running difference images showing the change in UV emission during the individual HXR bursts. Positive enhancements in the running difference images are overlaid on the HMI magnetograms and illustrate the apparent motion of the flare ribbons with respect to the magnetic field distribution.
In our analysis we specifically focus on three sub-regions that encompass the northern footpoints of the filament and flare loops (flare footpoint 1), the southern footpoint of the flare loops (flare footpoint 2), and the southern filament footpoint (filament footpoint). 

During HXR peak 1 at 11:21:26~UT, the filament was still within the FOV shown, and we can see traces of its eruption and connection to the southern filament footpoint in the base and running difference images. In the region of the northern footpoint (flare footpoint 1) several bright kernels are visible, which increased in brightness during the HXR burst. They are distributed along the major negative polarity region in the northern part of the AR. The brightest kernel probably corresponds to the footpoint of the flare loops, while some of the weaker kernels may have some connection to the erupting filament (see Fig. \ref{fig:AIAoverview}). In contrast to this compact northern flare ribbon, the southern flare ribbon already has an elongated shape during HXR peak 1 and covers the major positive polarity region in the south. However, the AIA images in Fig. \ref{fig:AIAoverview} show that the western part of this ribbon is not associated with the main flare loops. Instead, it corresponds to the smaller loop arcade that connects it to the minor negative polarity that lies parallel to this ribbon.

During the HXR peak 2 at 11:23:26~UT, the northern flare ribbon expands westward more closely matching the extent of the southern flare ribbon. As the flare progresses, a continuous westward motion of the northern flare ribbon can be observed in addition to the northward motion corresponding to the classical separation motion away from the PIL. This trend is best seen in the running difference maps and their contours. We see that the location of the 1600~Å enhancements during subsequent HXR bursts generally follows the shape of the major negative polarity. They pass close to the locations of the other bright kernels. For this reason, we have combined them into the same subregion for further analysis, as we are unable to separate their evolution over the course of the flare.

For the southern flare ribbon, the running difference images also show an apparent westward motion of the 1600~Å enhancements during subsequent HXR peaks, in addition to the southward motion away from the PIL. The combined motion takes them into the region between the major positive polarity region and the sunspot in the south of the AR. During HXR burst 6 a weak enhancement in this region is still visible. In addition, we see a stronger enhancement from the area of the southern filament footpoint, which appears as a discontinuity in the otherwise continuous evolution of the ribbon.

This evolution of the UV ribbons seems to be consistent with the STIX observations discussed in section \ref{sec:results_stix}. There we found that the northern HXR flare footpoint starts out relatively diffuse compared to the compact southern flare footpoint, then gradually brightens and becomes more compact until its appearance almost matches the southern flare footpoint. From the STIX perspective, the motion away from the PIL was observed. A slight westward drift might also be visible during the HXR peaks 1, 2, and 3, which had the best count statistics and therefore the highest accuracy for deriving the source position. However, the position of the flare, very close to the limb for STIX, makes it difficult to clearly identify such a westward drift. During the final HXR peak 6, STIX then observes the abrupt change of the southern source location.

To further verify the correlation between the non-thermal X-ray and UV observations, we examine the time evolution of the AIA 1600~Å counts integrated over the three sub-regions in Fig. \ref{fig:1600timeseries} and compare them with the STIX HXR flux evolution. We also include the counts for the whole 1600~Å submap shown in Fig. \ref{fig:1600images} to visualize the relative contribution of each subregion to the total brightness. In addition, we show the time evolution of the normalized total counts in each region, scaled to the minimum and maximum counts within the time interval plotted, to better illustrate the relative changes in each subregion.

We find a clear correlation between all major STIX HXR peaks (1 through 6) and the total AIA 1600~Å counts of the studied subregions. The three strongest HXR peaks (1, 2, 3) and the following two smaller bursts (4 and 5) clearly correlate with the 1600~Å counts from the subregions encompassing the flare ribbons, while the counts from the southern filament footpoint do not show an instantaneous response to these HXR peaks. Instead, they show a more gradual response that already starts during the onset of the filament eruption (i and ii) and later peaks around the time of HXR peak 2 before decaying again with the rest of the subregions.
We find that the same region around the southern filament footpoint shows another continuous increase in counts during the group of two smaller bursts (4 and 5), but peaks only afterward.

The last STIX HXR peak (6) is the important exception in its correlation with the 1600~Å counts in the studied subareas. We detect a significant relative increase in counts from the filament footpoint region during this time. Counts from the flare footpoint 1 sub-region also show a correlated but much weaker response. We find no response from the southern flare footpoint during this peak. These results further confirm the STIX observations, that show non-thermal emission mainly from the region of the southern filament footpoint during HXR peak 6, with a possible much weaker source at the northern footpoint.

\subsection{AIA differential emission measure analysis}\label{sec:results_dem}

\begin{figure*}
\centering
  \includegraphics[width=18cm]{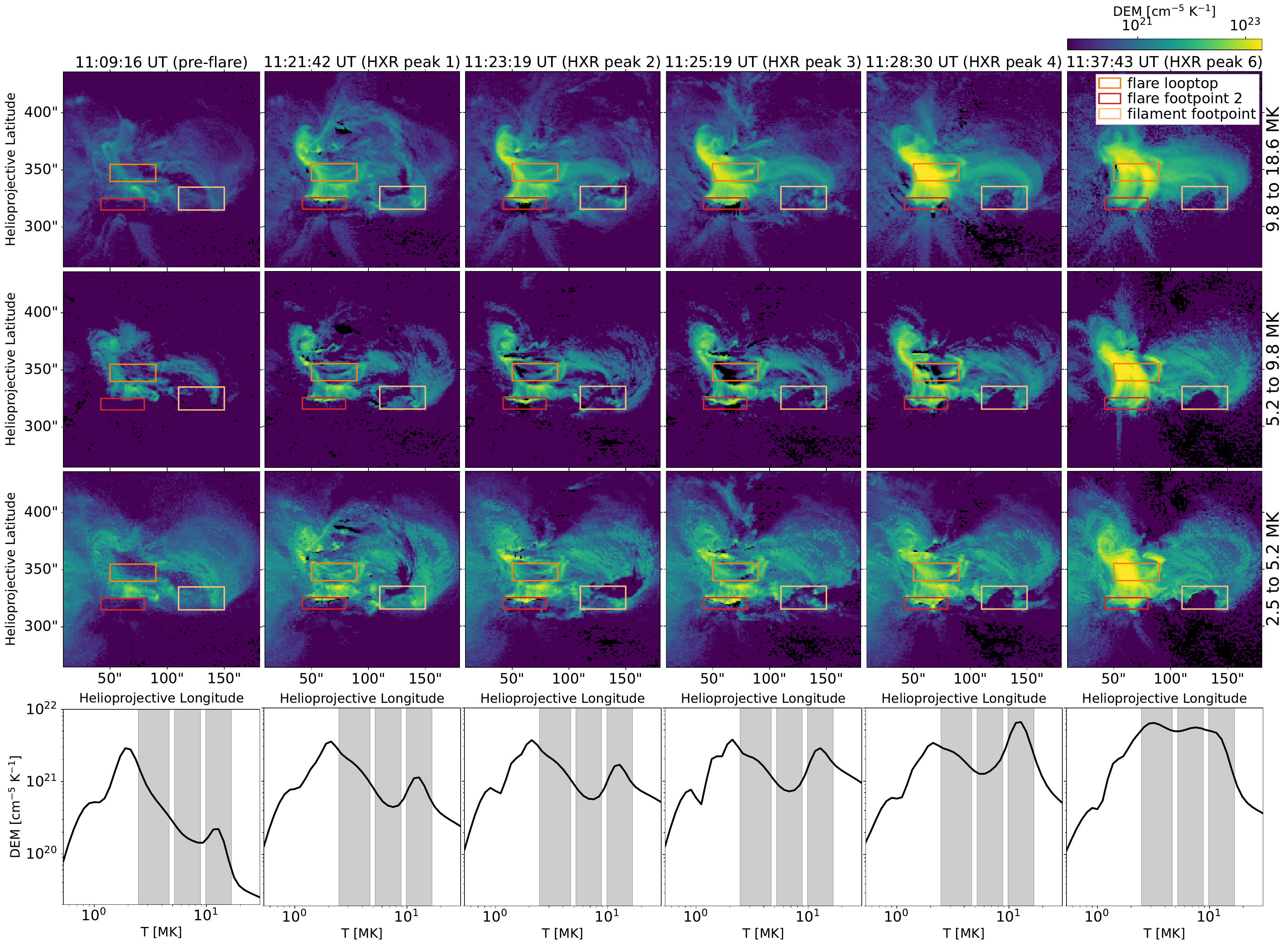}
    \caption{DEM reconstructions from selected times during the event.
    Top rows: DEM maps for six selected times (columns), showing for each pixel the mean DEM in three adjacent temperature ranges (rows): 9.8 -- 18.6 MK (first row), 5.2 -- 9.8 MK (second row), 2.5 -- 5.2 MK (bottom row). The time of the DEM reconstruction and the corresponding HXR peak are given at the top of each column.
    Bottom row: Mean DEM distribution of all pixels as a function of temperature for the selected times. The gray areas mark the temperature ranges for the DEM maps shown above.
    Movie: \href{https://drive.google.com/drive/folders/1ZiBDdO1Lk6x_q6X_3Y7LFT8V7NyQQQ3H?usp=sharing}{Google Drive}
    }
    \label{fig:DEMoverview}
\end{figure*}

\begin{figure}
  \resizebox{\hsize}{!}{\includegraphics{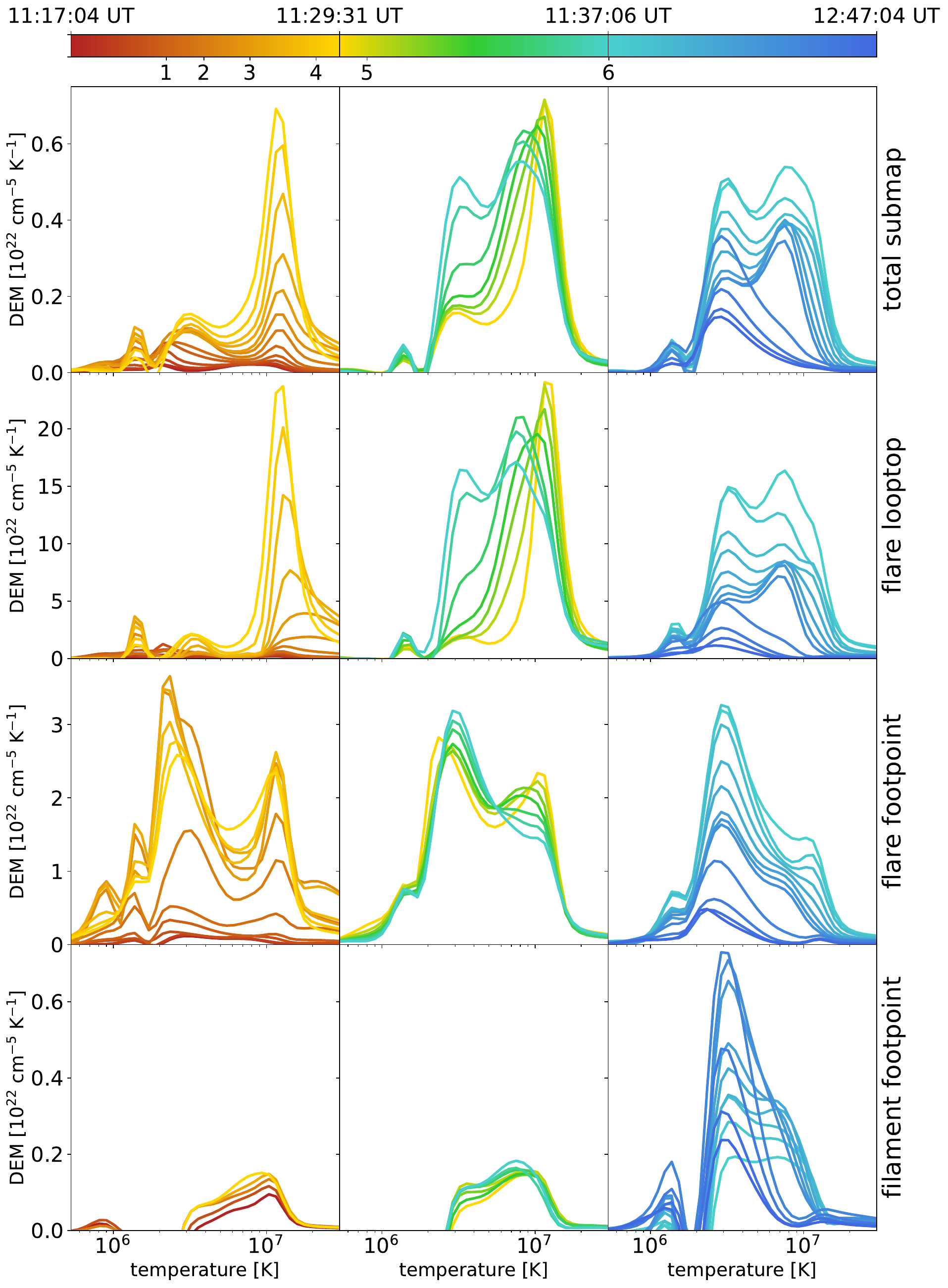}}
  \caption{Mean DEM distribution during three distinct phases (columns). Each row corresponds to one of the three subregions defined in Fig. \ref{fig:DEMoverview} or the whole FOV and is labeled on the right. Colors represent time, with marked time intervals having different durations. The times of the HXR peaks are marked at the bottom of the color bar.}
  \label{fig:DEMcurves}
\end{figure}

\begin{figure}
  \resizebox{\hsize}{!}{\includegraphics{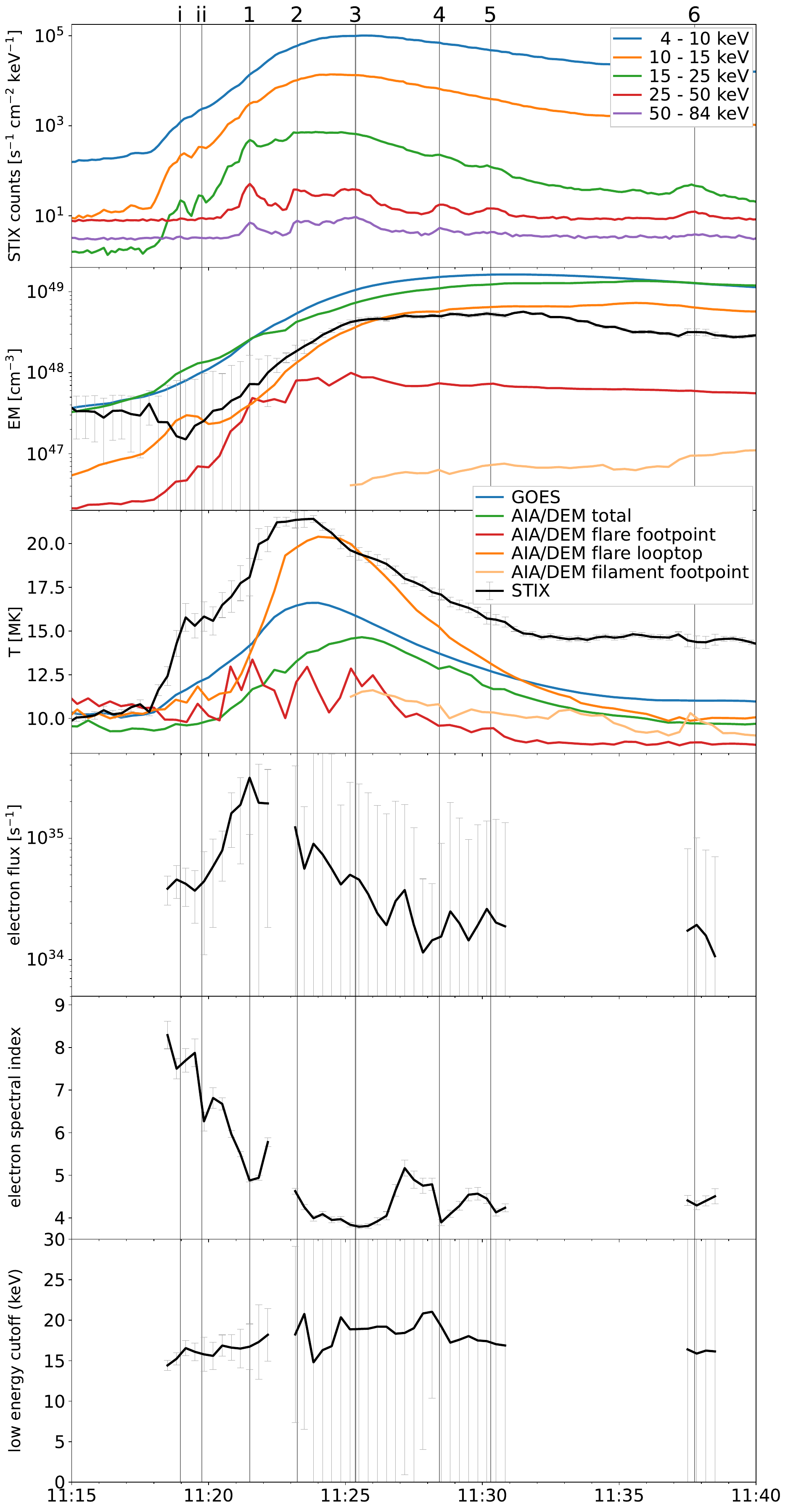}}
  \caption{Thermal and non-thermal emission characteristics from AIA, GOES, and STIX. From top to bottom: 
  1) STIX count flux in five energy ranges. Vertical lines mark the times of HXR peaks and are labeled on top. 
  2) Time evolution of the EM derived from the different instruments and for different subregions.
  3) Same as 2) but for the plasma temperature. 
  4) Electron flux with errors derived from the non-thermal fit to the STIX spectrum.
  5) Electron spectral index.
  6) Low energy cutoff of the accelerated electron distribution.
  }
  \label{fig:DEMtimeseries}
\end{figure}

Fig. \ref{fig:DEMoverview} shows an overview of the DEM analysis we performed to further study the characteristics of the heated flare plasma as it fills the coronal flare loops. We include DEM maps for three distinct temperature ranges. STIX should generally be sensitive to the hottest plasma shown in the top row. We show six selected times that match the times previously presented as part of the STIX overview in Fig. \ref{fig:STIXoverview}. The earliest time shows a pre-flare phase, where the filament is slowly rising, while the other five correspond to times during individual HXR peaks. We focus here on the gradual plasma evolution over this impulsive and main event phase because immediate changes to the DEM maps are often hard to track visually. We later use the EM and temperature evolution calculated for each of the subregions to better locate and correlate the plasma evolution with the energy input characterized by HXR emission.

The rising filament can be seen in the first column (11:09:16~UT) as an elongated dark structure. It is surrounded by the hotter plasma of the active region, which emits mostly in the 2.5 to 5.2 MK temperature range. The small loop arcade has a clear hot component, confirming that STIX saw it as the main thermal X-ray source during the pre-eruptive phase (see Fig. \ref{fig:STIXoverview}, first column).
Some hot loops extending over the northern part of the filament can also be seen in the 9.8 to 18.6 MK map. 
During the impulsive phase of the flare, the highest temperature range is quickly dominated by emission from the loop tops, consistent with the location of the major low-energy (thermal) STIX emission. They continue to increase in brightness until 11:28:30~UT before we see them cool down and appear in the lower temperature ranges.
On the contrary, the flare and filament footpoints are generally better seen in the lower temperature bands until the loops start to cool down.
Two slightly separated hot loop top sources are visible in the last 9.8 to 18.6 MK DEM map that could correspond to the double thermal structure discussed in the STIX results (section \ref{sec:results_stix}). The additional loops connecting from the main flaring region towards the southern filament footpoint continuously brighten in the highest temperature range throughout the flare.

In Fig. \ref{fig:DEMcurves}, we present the time evolution of the background-subtracted mean DEMs from three sub-regions marked in Fig. \ref{fig:DEMoverview}. We find that the DEM evolution is characterized by three distinct phases, visualized here by different parts of the color map. First, we see a phase of impulsive plasma heating over about 12 minutes during the filament eruption (i, ii), and the major energy release as indicated by HXR peaks 1 through 5. For the total map, we find that the DEM distribution rapidly rises with the main peak at >10 MK and a smaller one at about 3 MK. The DEM initially starts at even higher temperatures for the loop top but then quickly cools and peaks at roughly the same temperature as the total submap. In the DEM distribution of the southern flare ribbon, we find that a cooler 2 -- 3 MK component dominates this phase and develops a very sharp 2 MK peak. The DEM distribution around the southern filament footpoint only increases slightly during this phase.

The second phase is characterized by transitioning the hot >10 MK peak to a double peak DEM profile in the total map and the subregion of the loop top over eight minutes. This time interval approximately corresponds to the period between HXR peaks 5 and 6. The dominant >10 MK peak cools to lower temperatures, about 8MK, while the emission in the 3MK peak rises. The DEM distributions derived for the flare ribbons and the filament footpoints stay mostly the same.

During the last phase, we find a gradual cooling of the flaring plasma that we track for 1 hour and 16 minutes. The DEM decreases more rapidly at higher temperatures, leading to the transition to a single peak DEM distribution towards the end. This remaining peak moves to lower and lower temperatures as the overall DEM distribution decreases. The only exception to this trend is the DEM distribution derived for the southern filament footpoint area, which first increases in the 3 -- 4 MK temperature range before it also starts to cool gradually.

In Figure \ref{fig:DEMtimeseries}, we show the time evolution of EM and temperature obtained for different subregions and compare them with the values derived from GOES observations and from the isothermal fit to the measured STIX spectra. Further, we include parameters of the fitted non-thermal electron population for periods when we observed clear non-thermal emission in the STIX X-ray spectra. The values of the EM and temperature time evolution vary as expected between the three instruments due to their different temperature responses. However, the overall evolution agrees well, considering these differences.

The total EM derived from AIA shows a similar evolution to GOES. It misses some EM during the GOES peak of $1.6 \times 10^{49}$ cm$^{-3}$ at 11:31:17~UT but continues rising until around HXR peak 6, at 11:36:06~UT, with an EM of $1.35 \times 10^{49}$ cm$^{-3}$. Afterward, it surpasses the already decreasing EM observed by GOES. The AIA-derived temperature is lower throughout the flare and reaches its maximum of 14.6 MK at 11:25:43~UT after the GOES maximum of 16.2 MK at 11:24:02~UT. This behavior is expected since AIA's sensitivity extends to emission from much cooler plasma. The EM-weighted temperature will therefore be lower since there are large EM contributions from cooler plasma, and the total EM peaks later when all of the heated plasma has moved into the temperature range where AIA is more sensitive.
On the other hand, STIX only detects emissions from the highest temperature plasma (>10 MK), which lowers the observed EM compared to GOES and increases the derived plasma temperature. STIX's EM peaks at $0.6 \times 10^{49}$ cm$^{-3}$ at 11:31:29~UT, matching the peak time of GOES but with only 37.5 \% of its measured EM. The STIX temperature peaks at 21.4 MK at 11:23:49~UT, again almost matching the time of the GOES peak but with a peak temperature higher by about 5 MK.
Future efforts are needed to connect the STIX and AIA observations to derive a unified DEM distribution that is well-constrained over the whole temperature range present during a solar flare. This has already been successfully demonstrated for the RHESSI instrument \citep{Battaglia2015,Inglis2014}, and it would be very beneficial for future multi-instrument studies if such capabilities were extended to STIX.

In addition to our analysis of the total EM and temperature estimates for the whole AIA submap, we also derived these parameters for the plasma contained within the subregions marked in Fig. \ref{fig:DEMoverview}. We find that the total EM and mean temperature vary significantly between different regions. The EM from individual subregions shows us their relative contribution to the total submap's EM. 

The EM from the loop top source contributes the most to the total EM starting at HXR peak 1. It quickly rises and then follows the evolution of the total EM but with a lower overall value ($0.7 \times 10^{49}$ cm$^{-3}$ at its peak at 11:35:43~UT) that more closely matches the EM seen by STIX. The derived AIA/DEM temperature from the loop top also shifts much closer to the STIX observations in peak temperature (20.4 MK) and time (11:24:06~UT). This is consistent with the STIX imaging that shows the thermal plasma X-ray emission predominantly comes from the flare loop top (cf Fig. \ref{fig:STIXoverview}). The EM and temperature evolution in this subregion during the early phases of the event is heavily influenced by the motion and eruption of the filament. However, the filament quickly moves out of this subregion once it erupts, and from HXR peak 1 onwards, the EM and T evolution only reflects flaring plasma.

The EM from the associated flare ribbon shows an impulsive rise from the filament eruption until HXR peak 1. It continues to rise throughout HXR bursts 2 and 3, peaking at $0.1 \times 10^{49}$ cm$^{-3}$ at 11:25:19~UT, and then slowly decreases. It is also characterized by local EM peaks that correlate well with HXR peaks 1 to 5 and even seem to correlate with mainly thermal STIX peaks during the filament eruption (i, ii). The derived plasma temperature also correlates with the three main HXR peaks (1, 2, and 3), but the correlation with the other peaks is less pronounced. This correlation to the HXR peaks, in contrast to the smooth evolution observed for the loop top source, is a well-documented phenomenon. Both are generally related by the Neupert effect \citep{Neupert1968}, which is a strong indicator of energy transport by chromospheric evaporation and was originally documented as a multi-wavelength effect \citep[e.g.,][]{Dennis1993,Veronig2005}. These observations nicely demonstrate how traces of this phenomenon are present in EUV observations of plasma emission along the flare loops.

We only reconstructed the plasma parameters for the filament footpoint starting with HXR peak 3, since by this time the optically thick plasma from the erupting filament had completely moved out of this subregion. In this region, we find a continuous increase in EM throughout the shown time frame, with a sharper increase during and after HXR peak 6. The temperature decreases throughout the time evolution shown but shows a local maximum during the HXR peak 6.

From the non-thermal fit parameters, only the electron spectral index $\delta$ can be well-constrained throughout the flare. At the onset of the filament eruption, we find a spectral index of $\delta \approx 8$. The non-thermal spectrum hardens throughout the filament eruption (i, ii) and HXR peaks 1 and 2 until it reaches a maximum hardness of $\delta=3.8$ during HXR peak 3 at 11:25:29~UT. It subsequently softens slightly again through peaks 4, 5, and 6 but remains rather flat with a spectral index of $\delta = 4.5$. We find a clear soft-hard-soft behavior for all HXR peaks except number 2, as immediately after peak 1, the count rate strongly dropped, and the spectrum steepened so that we could not confidently fit a non-thermal model. With the onset of the HXR peak 2, the spectrum quickly hardens again, allowing us to fit the non-thermal model. However, the time resolution of 20 s is too low to accurately determine the spectral index at the onset of this HXR peak.

As soon as the non-thermal spectrum hardens and the EM and temperature of the thermal distribution rise, the lower energy cutoff is no longer well constrained because its counts are buried several orders of magnitude below the counts from the thermal X-ray emission. Our derived low energy cutoff stays mostly in the 15 to 20 keV range but contains large errors that, in turn, increase the uncertainty of the derived electron flux. We find a maximum electron flux of $3.15 \times 10^{35}$ s$^{-1}$  during HXR peak 1 that subsequently decreases during the remaining flare.

\subsection{Nonlinear force-free magnetic field extrapolation}\label{sec:results_magfield}

\begin{figure*}
\centering
  \includegraphics[width=18cm]{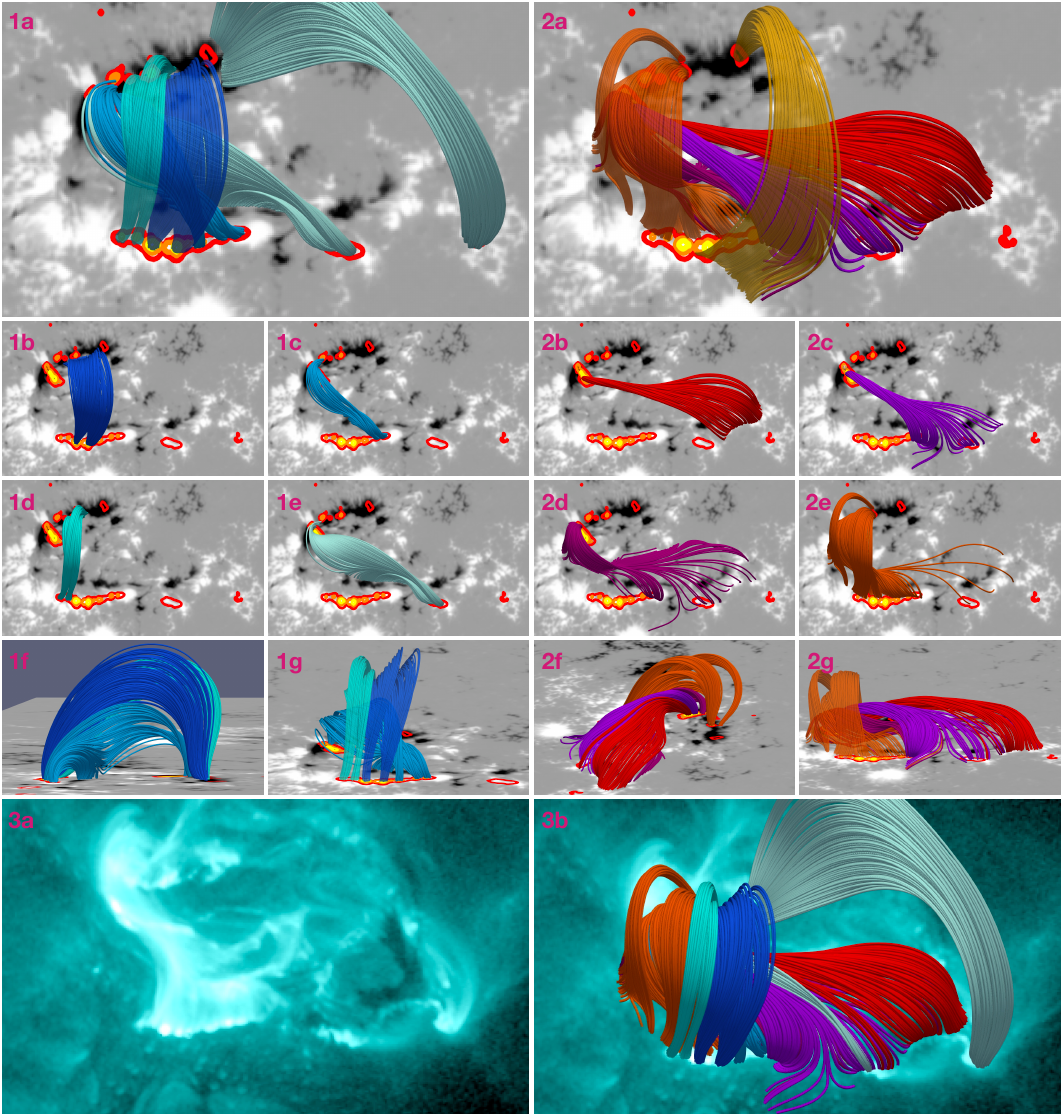}
    \caption{Field lines from a pre-flare magnetic field reconstruction (11:00~UT) that are associated with the AIA 1600~Å flare ribbon enhancements during HXR peak 1.
    Panel 1a: All field lines with seed points within a positive polarity region. 
    Panels 1b to 1e: Individual display of selected field lines. 
    Panels 1f and 1g: Combined field lines shown from two different perspectives.
    Panels 2a to 2g: The same structure but for field lines with seed points in negative polarity regions.
    Panel 3a: An AIA 131~Å image taken during HXR peak 1.
    Panel 3b: The combined structure of field lines with seed points selected from footpoints in both polarities.}
    \label{fig:magfield_overview}
\end{figure*}

\begin{figure}
  \resizebox{\hsize}{!}{\includegraphics{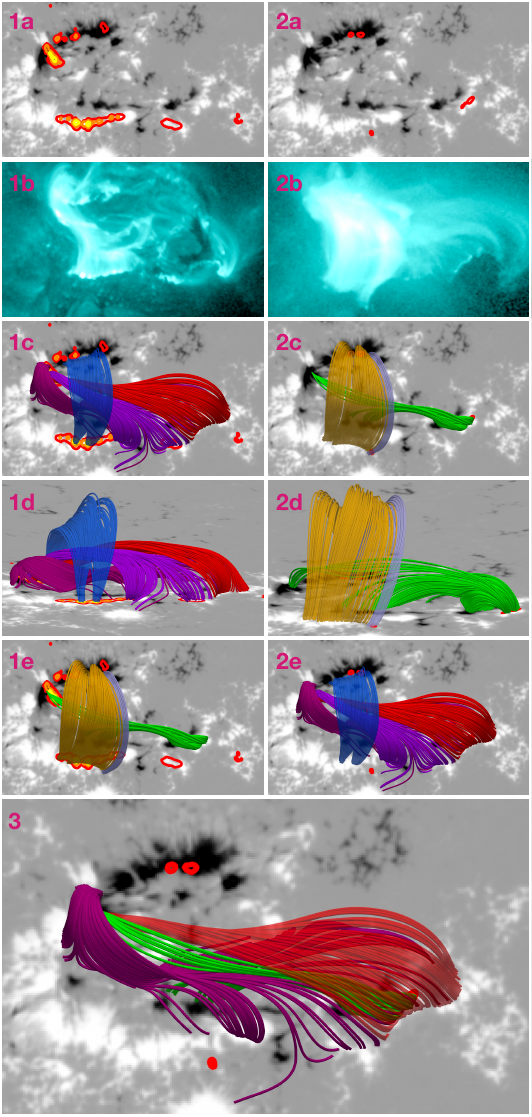}}
  \caption{Comparison of pre-flare magnetic field structures involved in the first (1) and last (6) HXR peaks.
  1a and 2a: HMI LOS magnetic field with contours of AIA 1600~Å enhancements during HXR peak 1 (left) and 6 (right). See Fig. \ref{fig:1600images}, bottom panels for the intermediate HXR peaks.
  1b and 2b: AIA 131~Å images taken during the two HXR peaks.
  1c and 2c: Selection of the most important field lines.
  1d and 2d: Side view of the same structure.
  1e and 2e: Same contours, but the field lines have been swapped.
  3: Comparison between selected field structures.
  }
  \label{fig:magfield_comparison}
\end{figure}

\begin{figure}
  \resizebox{\hsize}{!}{\includegraphics{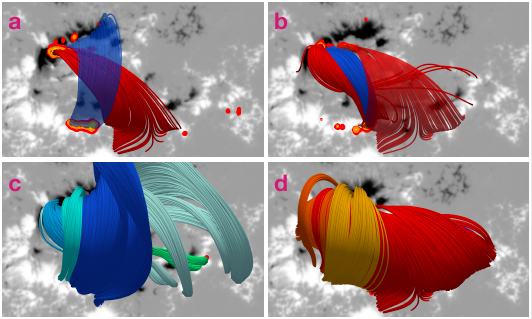}}
  \caption{Pre-flare magnetic field structures involved in the remaining HXR peaks. Panels a and b show the main fields that became involved in HXR peaks 2 and 3, respectively. Panels c and d show all field lines seeded during any HXR peak in the positive and negative polarity, respectively. Field lines from the first (1) and last (6) HXR peak are colored the same way as shown in Fig. \ref{fig:magfield_overview} while those from the HXR peaks in between are shown as blue and red.}
  \label{fig:magfield_other}
\end{figure}

Figure \ref{fig:magfield_overview} gives an overview of structures in the pre-flare NLFF magnetic field extrapolation associated with AIA 1600~Å enhancements during HXR peak 1. We use the 1600~Å enhancements outlining flare ribbons instead of the STIX non-thermal contours to avoid reprojection effects (from the Solar Orbiter to Earth view) and because of their superior spatial resolution. Our method is based on the assumption that the major 1600~Å enhancements are signatures of accelerated electrons moving toward the lower atmosphere along newly reconnected field lines, forming the bright flare ribbons. This assumption is supported by the good temporal correlation between STIX and AIA 1600~Å counts in different sub-regions shown in section \ref{sec:results_1600}. In addition, newly brightened flare loops observed by AIA follow the same apparent westward motion as was identified for the 1600~Å footpoints. We interpret this as a further indication that the location of the non-thermal footpoints likely matches that of the 1600~Å enhancements in the present event. We expect that the pre-event magnetic structures associated with the 1600~Å enhancements during the different HXR peaks became involved and activated in the reconnection process at that time.

In the following, we refer to the different polarity regions in the same way as introduced in section \ref{sec:results_aia} and as labeled in Fig. \ref{fig:AIAoverview}. There we have divided the part of the AR relevant to the present flare into two major polarity regions associated with the footpoints of the main flare loops. Both the major negative polarity region in the north and the major positive polarity region in the south are elongated structures that together form a U-shape. The major positive polarity region is also associated with the positive footpoints of the minor flare loop arcade. Immediately to the north and parallel to this major positive polarity region is the minor negative polarity region associated with the negative footpoints of the small flare loop arcade. The final relevant polarity region is the positive polarity region near the southern filament footpoint in the southwest of the AR.

Figure \ref{fig:magfield_overview} aims to give a broad overview of all relevant pre-flare magnetic field structures of the source AR that are associated with HXR peak 1 by dividing them into two groups according to the magnetic polarity of the 1600~Å contours in which their field line start points were placed. Panel 1a collectively shows all field structures associated with start points placed in a positive polarity region while panel 2a shows those with start points placed in a negative polarity region. We will discuss both groups individually before attempting to explain some of our observations in terms of their interactions.

Panel 1b shows the loops associated with the two 1600~Å maxima at the center of the flare ribbon in the major positive polarity region, while panels 1c and 1d depict field lines further out towards each end of the ribbon. They reveal how the magnetic field transitions from a loop arcade to field lines extending beneath this overlying structure. The field lines in panel 1c also partially connect to the minor negative polarity region.
In panels 1f and 1g, these field lines (1b, 1c, 1d) are shown collectively from two additional perspectives, highlighting the twist and the transition in their connection from the major to the minor negative polarity region. 
Panel 1e shows field lines seeded in the positive polarity region near the southern filament footpoint. Their connection also transitions from the minor to the major negative polarity region, where they connect to the main negative flare kernel. Field lines close to the transition between these two connectivities are strongly curved and extend toward the minor negative polarity kernels. In panel 1a, one additional large loop structure is shown that starts in the positive polarity region around the filament footpoint and extends toward the western end of the major negative polarity.

Panel 2b shows the magnetic structures derived from seed points in the 1600~Å maxima in the negative polarity. The field lines start out as a twisted magnetic structure but eventually spread out into a sheet that partially runs along the boundary of the semicircular positive polarity region associated with the southern filament footpoint. 
The field lines in panel 2c start from the same major flare kernel. We again see a twisted magnetic structure that crosses over the field lines in panel 2b and then spreads out into a sheet that transitions from the western end of the positive polarity flare ribbon to part of the positive polarity filament footpoint area.
Panel 2d shows field lines from the third group of field line seed points placed in the major flare kernel. These field lines first pass over the previous field structures, but then curve around them and fill the space underneath. The result are field lines with connections to both the major positive polarity ribbon and the southern filament footpoint. The latter ones are part of the filament structure and lie beneath all other magnetic field structures involved in HXR peak 1. 
Panel 2e shows the field lines seeded in the two minor kernels along the major negative polarity region. Field lines with their seed sources located farthest from the major flare kernel are part of the loop arcade and connect to the eastern end of the major positive polarity region. Those with seed sources closer to the major kernel curve below this loop arcade toward the western side of the positive polarity ribbon. Some field lines extend even further toward the southern filament footpoint. Therefore, parts of the minor negative polarity kernels (at least the one closest to the major kernel) appear to possess somewhat connection to the filament.
Field lines associated with the westernmost kernel of this major negative polarity ribbon are only shown in panel 2a and are part of a large overlying loop structure. Panels 2f and 2g show two different perspectives of the combined field line structure that further highlights the twist in field structures close to the filament. We see how the flux holding filament, visualized by those field lines extending from the major negative kernel to the southern filament footpoint in 2d and 2e, generally lies beneath all other structures involved in HXR peak 1. One can imagine how this filament structure erupted from its pre-event state shown here, stretching the overlying field, and forming the reconnection region below.

In panels 3a and 3b we show the relation of the pre-flare field lines with seed points in both polarities to the AIA 131 Å observations at the time of HXR peak 1. We expect that bright loops in AIA are generally the result of reconnection between the field structures shown.
The main flare loops extend between the maxima of the 1600~Å enhancements in both polarities. This suggests that they are mainly the result of reconnection between the blue field lines of panel 1b and the red and purple field lines of panels 2b and 2c, respectively. Their possible reconnection is described in more detail in Fig. \ref{fig:magfield_comparison} and its discussion.
The small loop arcade appears to be the result of an interaction between some of the field lines that connect to the minor negative polarity and the western region of the major positive ribbon. Structures with components of possible candidates are emphasized in panels 1c, 1e, 2d, and 2e.

Bright loops extending northward from the main flare region toward the minor flare kernels may be related to some loops of the overlying arcade slipping off and being reconnected during the filament eruption. Such slipping of overlying loops can be seen in the AIA movies leading up to this HXR peak.
The large loops starting west of the southern filament footpoint (panel 1a) may play a role in the diffuse S-shaped hot loops visible in AIA 131 and 94 Å (e.g., Fig. \ref{fig:AIAoverview}, 11:43:59 UT) that appear to connect from this region towards the main flaring region in the negative polarity. Similar field structures associated with other HXR peaks are displayed in Fig. \ref{fig:magfield_other}, and together give some idea of the field structures involved in creating this loop system to the west of the flaring region.

Figure \ref{fig:magfield_comparison} focuses on the field lines that may be involved in the main reconnection during HXR peak 1 (panels 1a to 1e) and compares it with the HXR peak 6 (panels 2a to 2e), where we observed the jump in the HXR footpoint location. HXR peak 1 seems to be mainly a result of the reconnection between the blue and red/purple field lines, driven by the eruption of the underlying filament structure. Part of the original filament structure also seems to have been involved in the reconnection. The reconnection between the blue and red/purple field structures formed the narrow flare loop arcade between the strongest 1600~Å contours in each polarity that we observe in AIA 131~Å. The negative foopoints of the blue field lines lie very near the corresponding flare kernels and are also consistent with an involvement of this flux in the flare reconnection if a small error in the NLFF field is allowed for. The positive footpoints of the red/purple field lines spread out over a larger area, so that the formation of flare kernels at these locations is less likely.

1600~Å enhancements from subsequent HXR peaks (see bottom panels of Fig. \ref{fig:1600images}) follow the path of the blue field lines in the negative polarity region. In the positive polarity region, subsequent 1600~Å enhancements drift southwest in a direction that does not correspond to the field lines reconnected during HXR peak 1. Only during HXR peak 6 does the reconnection involve field lines connected to the southern filament footpoint.

Comparing the field lines shown for HXR peak 6 (Fig \ref{fig:magfield_comparison}, panels 2c and 2d) with 1600~Å enhancements of previous HXR peaks (e.g., panel 1e), it is clear that we should not expect them to still be connected as shown.
Field lines starting in the major negative polarity region (yellow) connected to a region mostly covered by the flare ribbon during HXR burst 1, with the rest lying in the path of subsequent HXR bursts. Fields extending from the positive polarity around the southern filament footpoint (green) were originally connected to the major negative flare kernel during HXR peak 1. Consequently, both pre-event field structures associated with HXR peak 6 have already been involved in reconnection and will have had a different topology at the time of this late HXR peak.

In panel 3 of Fig. \ref{fig:magfield_comparison} we compare the green field lines of HXR peak 6 with those of HXR peak 1. It is clear that they are closely related to the erupting filament structure, as they lie completely below the red field lines that surround the filament. In conclusion, the HXR footpoint at the southern filament footpoint is likely the result of reconnection involving field lines that were closely related to the erupting filament and that were already involved in earlier reconnection episodes.

In Fig. \ref{fig:magfield_other} we show the pre-flare field lines associated with the remaining HXR bursts. Panels a and b present the main flux systems associated with HXR peaks 2 and 3, respectively. For these, we found similar counts in STIX HXR and AIA 1600~Å compared to HXR peak 1. For HXR peak 2, field lines seeded in the negative polarity still extend below those in the positive polarity. Together they form an X-like structure with the eastern footpoints closer to the PIL and associated with the reconnection. During HXR peak 3, the associated flux systems were closer to a single sheared arcade.
Panels c and d show fields seeded in the positive and negative polarity, respectively, during any of the six HXR peaks.
In panel c, we see that the flux rooted in the positive polarity ribbon generally reveals higher and larger loop structures as seed points located further westward are considered. Comparison with the 1600~Å footpoints in Fig. \ref{fig:1600images} shows that the larger loops near the sunspot were reconnected mainly during HXR peak 4 at 11:28~UT, which is associated with a rapid expansion of the ribbon toward the sunspot. This rapid southward expansion of the ribbon corresponds to the time when we observed the double thermal source with STIX (see Fig. \ref{fig:STIXoverview}, column 5). We also found a matching hot double loop structure in the DEM reconstructions (see Fig. \ref{fig:DEMoverview}, top right panel), where one of the loop arcades seems to extend towards the sunspot.

In addition, panel c in Fig. \ref{fig:magfield_other} shows more field lines starting in the region of the southern filament footpoint (light blue). Reconnection involving these field lines is likely to result in the diffuse S-shaped hot loop structure extending from the main flare region towards the southern filament footpoint. Their evolution during the flare was characterized by the gradual response in AIA 1600~Å to all but the last HXR burst (number 6) (see Fig. \ref{fig:1600timeseries}, bottom panel) and by DEM reconstructions showing that these loops are gradually heated to high temperatures as the flare progresses (see Fig. \ref{fig:DEMoverview}, top row). Since 1600~Å emission in this area is generally not correlated with the HXR bursts, we should expect the field lines shown to be a random selection of the fields active during the individual HXR burst and not necessarily the most important field lines for the more continuous reconnection in the area.
Panel d shows the combined field structure starting from the negative ribbon. It is much more compact compared to the blue field lines in panel c and wraps tightly around the entire filament structure, covering both the positive major ribbon and the fields connecting towards the southern filament footpoint.

\section{Discussion}\label{sec:discussion}

We presented an analysis of the energy release, transport, and particle acceleration in the M4-class flare on March 28, 2022, demonstrating the advantages of STIX imaging and spectroscopy as part of a multi-instrument flare study that uses multiple perspectives to constrain observations. Solar Orbiter was at a large separation angle of $83.5^\circ$ west of the Sun-Earth line and near its first perihelion during the nominal science phase, at a distance of only 0.33 AU from the Sun. This resulted in a vastly different view of the event under study compared to the near-Earth observations from SDO. For Solar Orbiter, the flare appeared close to the east solar limb, giving us a side-on view of the event that provides a clear separation between non-thermal footpoints and thermal loop top sources. At the same time, the event was seen on the solar disk by AIA. This top-down view minimizes distortions of chromospheric sources for ideal spatial separation of flare ribbons and filament footpoints that can be related to STIX non-thermal sources. In addition, high-cadence, high-resolution AIA EUV observations are available to constrain the locations of STIX thermal sources along the line of sight. The flare location is also advantageous for the magnetic field extrapolation performed. Our main goal for this study was to take advantage of this instrument configuration to analyze the signatures of energy release and transport by accelerated electrons and how they relate to the erupting filament and the magnetic field configuration.

The reprojections of the STIX nonthermal sources and the line-of-sights through the thermal STIX sources to the AIA images revealed the evolution of two non-thermal footpoint sources and thermal emission from the flare loops evolving below the erupting filament (see Fig. \ref{fig:STIXoverview}). The inferred locations and evolution of the X-ray sources closely match the AIA EUV observations and the general expectations of an eruptive flare, but show some interesting deviations. Most notably, we find that during the late HXR peak number 6, the non-thermal X-ray emission mostly originated from the southern anchor point of the erupting filament in contrast to the two classical X-ray footpoints located at the flare ribbons during previous HXR peaks. We also detect possible additional emissions from the northern flare footpoint, which is very close to the northern filament anchor point, as well as from a coronal source. The validity of these source reconstructions was demonstrated by a STIX imaging robustness analysis (see Fig. \ref{fig:perturbations_overview}) as well as through correlated AIA 1600~Å brightenings (see Fig. \ref{fig:1600images} and \ref{fig:1600timeseries}). The chromospheric UV emission closely correlates with the timing of all HXR bursts and supports the source locations of the flare and filament footpoints derived through STIX image reconstruction. During HXR peak 6, chromospheric emission from the edge of the southern filament anchor point agrees remarkably well with the reprojected non-thermal STIX contours. Even the more uncertain STIX observations of non-thermal emission from the northern flare and/or filament footpoint seem to be supported by a correlated AIA 1600~Å response.

\citet{Stiefel2023} recently reported the first observations of non-thermal emission from two filament footpoints detected by STIX during the impulsive phase of an eruptive M1.8 flare. This discovery followed observations from \citet{Chen2020}, who reported microwave emission from non-thermal electrons accelerated by the flare that travel along the filament. However, this phenomenon of flare-accelerated electrons accessing the filament was only observed during the early impulsive phase of the flare and coincided with stronger emission from the main flare footpoints. Both the microwave emission for the filament and non-thermal X-ray emission from its footpoints quickly faded as the flare progressed. \citet{Chen2020} attributed this quick fading of the microwave source to the increasing difficulty for accelerated electrons to access the filament due to the strong-to-weak shear transition of reconnecting field lines \citep{Aulanier2012}. Our observations of non-thermal emission from the location of the filament footpoints occur during a late HXR peak (number 6), about 16 minutes after the first HXR burst (number 1). This does not support the suggestion in \citet{Chen2020}, according to which this progressed state of the flare would make it very difficult for flare-accelerated electrons to access the filament and return to the chromosphere. Our observation is consistent with the filament becoming involved in the flare reconnection \citep{Gibson2008,Aulanier2019}.

Following the full evolution of the 1600~Å flare ribbons (see Fig. \ref{fig:1600images}, bottom row), we found a continuous westward drift during HXR bursts 1 to 5, in addition to their motion away from the PIL. During HXR peak 6 this westward trend continues. However, the evolution of the positive polarity 1600~Å flare ribbon shows a discontinuity as it suddenly jumps to the area of the southern filament footpoint. 
The same westward trend is seen in the EUV post-flare loops. The flare arcade is initially very narrow, extending only between the strongest 1600~Å kernels during HXR peak 1. The flare arcade then rapidly expands westward during subsequent HXR peaks. Eventually, we observe loops extending from the northern flare footpoint area towards the southern filament footpoint area, which appear just after HXR peak 6. They can be seen in the final two time steps included in Fig. \ref{fig:AIAoverview}. At 11:43:59~UT, five minutes after HXR peak 6, we observe hot flaring loops in the AIA 94 and 131~Å channels. These loops become more visible once the plasma has cooled enough to be detected in the 193 and 211~Å channels, as shown in the images taken at 12:10:47~UT. We again refer to the movie accompanying Fig. \ref{fig:AIAoverview} to see the full evolution and changes in the (post-) flare arcade systems.

We assume that the 1600~Å enhancements are signatures of the deceleration of non-thermal electrons and therefore correlate with HXR sources, which is supported by the correlation between the time series of STIX and AIA 1600~Å counts (see Fig. \ref{fig:1600timeseries}). Therefore, we should expect the non-thermal X-ray sources to follow their westward motion. However, due to the side-on view that STIX had of this event, the westward motion of the sources is not easily seen, and the footpoints appeared as standard X-ray sources moving away from the PIL. The late HXR peak number 6 shows a sudden jump in the southern footpoint location in very good agreement with the jump of the 1600 Å ribbon.

Spectral fitting of HXR peak 6 revealed parameters of the accelerated electron population consistent with the range of values derived during the previous HXR peaks and their evolution (see Fig. \ref{fig:DEMtimeseries}, bottom three panels). The electron spectral index of $\delta=4.5$ during HXR peak 6 continues the slight softening of the spectrum observed during HXR peaks 4 and 5 from its maximum hardness during HXR peak 3. In addition, we found a similar low energy cutoff and electron flux as during the previous HXR peaks. This continuous evolution of the derived parameters suggests a similar nonthermal electron population and consequently a similar acceleration process and transport conditions. Thus, it supports the idea of a continuous reconnection process during all HXR peaks as opposed to a sudden change during HXR peak 6.

A motion of the X-ray footpoints parallel to the inversion line has been previously reported in statistical studies by \citet{Bogachev2005} using the Hard X-ray Telescope on board Yohkoh and by \citet{Gan2008} using RHESSI. It is generally attributed to the shift of the acceleration region along the PIL. 
\citet{Liu2009} further demonstrated how this phenomenon could be related to asymmetric filament eruptions that erupt in a whhipping-like or zipping-like fashion as a response to an external asymmetric magnetic confinement. Indeed, the event under study may fit the characteristics of a whipping-like filament eruption, with its more active leg in the north whipping upward while the more anchored leg in the south remains fixed to the photosphere. \citet{Liu2009} argued that the X-ray sources would then be expected to shift towards the anchored leg due to the subsequent reconnection of loops in the overlying arcade, as observed in the present event.

Using a NLFF extrapolation of the pre-flare magnetic field based on a new physics-informed neural network method \citep{Jarolim2023}, we were able to investigate in more detail the relationship between the erupting filament and the field lines that likely got involved in the reconnection during the different HXR peaks.
We provided evidence that HXR peak 1 was the result of reconnection between strongly sheared magnetic field structures very close to, or possibly even part of, the erupting filament and weakly sheared loops from an overlying arcade (see Fig. \ref{fig:magfield_comparison}, panels 1c and 1d).
For HXR peak 6, we again found the involvement of field lines that were originally very close to the filament structure.

Our NLFF extrapolation combined with the multi-instrument analysis not only explains the evolution of the HXR sources but also sheds light on several other observations. It reveals a complex magnetic configuration involved in the flare, characterized by multiple interacting loop systems. This magnetic complexity could be related to many of the presented observations, such as the amplification of the small loop arcade, the extension of bright S-shaped hot loops away from the main flare region towards the southern filament footpoint, and the double thermal loop top source. Our results demonstrate how we can use multi-instrument multi-point remote sensing observations to obtain a much more complete understanding of the flare physics, especially in complex magnetic configurations.

At this point, we would like to highlight the potential of the Advanced Space-based Solar Observatory \citep[ASO-S;][]{Gan2019_ASO-S} launched in October 2022, with its Hard X-ray Imager \citep[HXI;][]{Zhang2019_HXI} that observes solar flares in the 30 -- 200 keV energy range from a sun-synchronous orbit. Combined with STIX observations, HXI will enable future spectroscopic observations of thermal and non-thermal X-ray sources, allowing us to better constrain their location and morphology, and to investigate the directional dependence of the measured X-ray spectra.

\section{Conclusion}\label{sec:conclusion}

The multi-instrument study of the March 28, 2022 M4-class eruptive flare, with observations from two vastly different vantage points, used STIX imaging and spectroscopy, EUI and AIA imaging, and a HMI NLFF magnetic field extrapolation to successfully constrain signatures of energy transport by accelerated electrons and show their connection to various magnetic structures, including the erupting filament.
Our analysis revealed the evolution of non-thermal footpoints and thermal loop sources that show some unexpected deviations from a simple two-ribbon flare. 

In particular, we found a shift of the southern non-thermal flare footpoint to the erupting filament's anchor point during a late HXR peak (number 6). These observations were confirmed by a robustness analysis of the STIX imaging and a correlated AIA 1600~Å response.
The full evolution of the AIA 1600~Å flare ribbons further indicated that the sudden jump in footpoint configuration behaved as a discontinuity in an otherwise continuous westward drift of the footpoints throughout the flare.
Using the NLFF magnetic field extrapolation, we showed that the shift of the HXR footpoints cannot be exclusively explained by subsequent reconnection along a pre-existing overlying arcade. Instead, we found that the footpoints during the first and last HXR peaks are associated with field structures that were initially very close to, or even part of, the erupting filament structure. These strongly sheared field lines probably reconnected with weakly sheared field lines of an overlying arcade during the first HXR peak and became part of the reconnection again during the last HXR peak.

In addition, we found a complex evolution of the STIX thermal X-ray sources, with multiple sources during the first HXR peak and a very low initial loop top source that later transitions through a double structure before re-emerging and rising as a single source. We verified these observations with DEM analysis and related them to the complex flare geometry with its multiple interacting loop systems using the NLFF magnetic field extrapolation.

Our observations highlight the complex nature of solar flares and show how existing models can be an oversimplification, in particular when dealing with more complex magnetic configurations. We have shown how reconnection, especially in the early phases of the flare, can involve field lines with very different shear, and how this is important to fully explain the footpoint motions throughout the flare. This generalizes a simpler model known as whipping filament eruption and reconnection, which only considers the propagation of reconnection along an arcade with uniform shear. Our results can aid in the interpretation of future studies of energy release and HXR footpoint motions, and demonstrate the importance of a multi-instrument, multi-point approach.

\begin{acknowledgements}
The authors thank Dr. Cooper Downs for insightful discussions on the magnetic connectivities in this event during the workshop of the International Space Science Institute (ISSI) team no. 516 on “Coronal Dimmings and their Relevance to the Physics of Solar and Stellar Coronal Mass Ejections.
Solar Orbiter is a space mission of international collaboration between ESA and NASA, operated by ESA.
The STIX instrument is an international collaboration between Switzerland, Poland, France, Czech Republic, Germany, Austria, Ireland, and Italy. 
The EUI instrument was built by CSL, IAS, MPS, MSSL/UCL, PMOD/WRC, ROB, LCF/IO with funding from the Belgian Federal Science Policy Office (BELSPO/PRODEX PEA 4000134088); the Centre National d’Etudes Spatiales (CNES); the UK Space Agency (UKSA); the Bundesministerium für Wirtschaft und Energie (BMWi) through the Deutsches Zentrum für Luft- und Raumfahrt (DLR); and the Swiss Space Office (SSO).
S.P., E.C.M.D, and A.M.V. acknowledge the Austrian Science Fund: project no. I 4555.
A.F.B. and S.K. acknowledge the Swiss National Science Foundation Grant 200021L\_189180 for STIX.
B.K. acknowledges support by the DFG and by NASA through Grants No. 80NSSC19K0082 and 80NSSC20K1274. 
\end{acknowledgements}

\bibliographystyle{aa} 
\bibliography{bib-file} 

\end{document}